\documentclass[12pt,epsf,letter]{report}
\usepackage{VD-dissertation}

\usepackage[phd]{thesisASTformat}
\degreetitle{
             Constraints on Compact Binary\\
             Formation and Effective\\
             Gravitational Wave Likelihood\\
             Approximation\\
            }
\degreeauthor{Vera Del Favero}
\degreedate{August 25, 2022}
\advisor{Dr.\ Richard O'Shaughnessy}
\memberA{Dr.\ Jason Nordhaus}
\memberB{Dr.\ Patricia Schmidt}
\comChair{Dr.\ Ernest Fokoue}
\ASTdirector{Dr.\ Andrew Robinson}

\usepackage[margin=3cm]{geometry}

\usepackage{fancyhdr}
\begin{document}

\clearpage{}\pagestyle{empty}

\begin{center}
\textbf{
\begin{huge}
    {\mytitle}
\end{huge}
}
\end{center}

\vspace{2.0cm}

\hrule
\begin{center}
\begin{Large}
    A dissertation submitted in partial fulfillment of the requirements \\
    for the degree of {\letters} \emph{\kindspelledout} \\
	in Astrophysical Sciences and Technology\\
	\textit{Vera Del Favero}\\
\end{Large}
\end{center}
\hrule

\vspace{2cm}

\vspace{3.8cm}

\begin{center}
\begin{Large}
    School of Physics and Astronomy \\
    Rochester Institute of Technology\\
    Rochester, New York \\
    August 25, 2022 \\
\end{Large}
\end{center}

\cleardoublepage
\clearpage{}

\newgeometry{left=3cm,top=3cm,bottom=3cm,right=3cm}
\setstretch{1.375}

\makecertificatepage

\newgeometry{left=3cm,top=3cm,bottom=3cm,right=3cm}

\maketitlepage

\clearpage{}\chapter*{Declaration}

\begin{doublespacing}

I, VERA E. DEL FAVERO (``the Author''), declare that no part of this dissertation is
    substantially the same as any that has been submitted for a degree
    or diploma at the Rochester Institute of Technology or any other 
    university.
I further declare that the work in chapter \ref{chap:popsyn} is entirely my own.
Much of \ref{chap:binary} consists of a review of the related literature,
    however the sections describing the properties of new
    population models are my own.
Chapters \ref{chap:intro}, \ref{chap:nal}, and \ref{chap:conclude}
    draw in parts on work done with collaborators, including contributions
    from other authors.
Those who have contributed scientific or other collaborative insights are
    fully credited in this dissertation,
    and all prior work upon which this dissertation builds is cited appropriately
    throughout the text.

Modified portions of this dissertation will be published by the author
    and her advisor, in a journal(s) yet to be determined.

\begin{itemize}
\item A substantial part of \textbf{Chapter \ref{chap:nal}} is based on the paper
    \cite{nal-chieff-paper},
    entitled \emph{Normal Approximate Likelihoods to Gravitational
    Wave Events}, authored by {V.} Delfavero, {R.} O'Shaughnessy,
    {D.} Wysocki, and {A.} Yelikar.
\item The remainder of \textbf{Chapter \ref{chap:nal}} is based on the paper
    \cite{nal-methods-paper},
    entitled \emph{Compressed Parametric and Non-Parametric Approximations
        to the Gravitational Wave Likelihood}, authored by {V.} Delfavero,
        {R.} O'Shaughnessy, {D.} Wysocki, and {A.} Yelikar.
\end{itemize}

\end{doublespacing}
\clearpage{}

\clearpage{}\begin{abstract}

Since the initial discovery of gravitational-waves from merging
    black holes,
the LIGO Scientific Collaboration together with Virgo and KAGRA 
    have published 90 gravitational-wave observations of compact binary mergers 
    in the Gravitational-Wave Transient Catalog papers.
One of the quintessential questions of this decade in
    gravitational-wave astronomy is the characterization
    and impact of the population of observed gravitational-wave
    signals from merging black holes and neutron stars.
Now, there is greater incentive than ever to study 
    the formation channels for these compact binary mergers.
In this work, we carry out an investigation of
    isolated binary evolution formation channel,
    comparing predictions of the gravitational-wave
    population from the StarTrack synthetic universe simulations
    to the observed population of compact binary mergers
    in order to constrain certain astrophysical processes in binary evolution.
In due course, we construct, apply, and provide parametric
    and non-parametric models for the likelihood function
    of the full set of astrophysical parameters of each event
    in the Gravitational-Wave Transient Catalogs,
    including truncated multivariate normal distributions
    normalized on a bounded interval,
    which we have shared in our associated publications
    \cite{nal-chieff-paper,nal-methods-paper}.
We present the findings of our investigation
    of the formation parameters for the isolated binary evolution
    formation channel for compact objects.
We have uncovered confounding systematic effects in our model
    by considering the agreement and disagreement
    of predictions based on the event rate and mass distribution.
Furthermore,
    our preliminary results demonstrate
    the benefits of a multi-dimensional analysis which 
    is sensitive to the interdependence of the predicted
    detection population on many formation parameters.
Our essential contribution is therefore
    a method for carrying out such an analysis efficiently,
    while considering its self-consistency.
We discuss potential sources of bias
    as we also present the properties of our best binary evolution models,
    which are consistent with unrestricted stellar mass loss due to winds,
    high mass and angular momentum loss to ejected portions of a 
    common envelope, and substantial black hole supernova recoil kicks.
We conclude with a discussion of the impact of these activities
    for gravitational-wave and multi-messenger astronomy.

\end{abstract}
\clearpage{}

\tableofcontents

\newgeometry{left=3cm,right=3cm,top=3cm,bottom=3cm}
\pagestyle{fancy}
\fancyhf{}
\cfoot{\thepage}

\begin{chapter}{Introduction}
\label{chap:intro}

On September 14th, 2015, the first gravitational-wave signal from
    a coalescing binary-black-hole was detected by the 
    Laser Interferometer Gravitational-Wave Observatory (LIGO),
    and reported by the LIGO Scientific Collaboration (LSC)
    \cite{GW150914-detection, GW150914-astro, GW150914-num}.
Now, after the third observing run of LIGO together with Virgo and KAGRA
    (KAmioka GRAvitational-wave detector),
    astronomers have reported a variety of
    compact binary merger events,
    including Binary Black-Hole (BBH), 
    Binary Neutron-Star (BNS), 
    and Neutron-Star Black-Hole 
    (NSBH) systems.
    \cite{O3-Detector, Virgo, Kagra}.
These observations have been summarized by many groups
    including the ninety events reported in the
    Gravitational-Wave Transient Catalogs (GWTCs)
    as well as independent analysis
    \cite{GWTC-1, GWTC-2, GWTC-3, GWTC-2p1,
        IAS, 3-OGC, Boyle_2019_SxS,
        Jani_2016_GA_Tech, BAM2008}.
In the era of multi-messenger astronomy,
    these and future gravitational-wave observations
    describe a view of the cosmos set by more than just
    electromagnetic radiation.

Each gravitational-wave signal is a strain imposed on the interferometer
    instruments.
These signals are identified by a rapid search,
    and events which may have electromagnetic counterparts 
    are immediately constrained by a low-latency parameter estimation
    pipeline
    \cite{Aasi_2014,Singer_2016, PhysRevLett.127.241103}.
Following this, all events are then 
    characterized by robust Bayesian statistical inference
    in gravitational-wave parameter estimation (PE)
    \cite{Pankow2015,Veitch2015LALInference, 
    2019PASP..131b4503B, RIFT, Bilby2019, IAS}.
Such an inference evaluates the agreement between
    strain information and relativistic waveform models,
    to yield accurate parameter estimates
    for each detection.

Once the properties of individual 
    gravitational-wave sources are characterized,
    a question which arises as to how these systems
    form in The Universe.
The formation of these compact binary objects
    can be described by a variety of channels,
    one of which is the formation of massive stellar binaries by isolated
    evolution
    \cite{1991ApJNarayan, Belczynski2002, Belczynski2008,
        Belczynski2020, broekgaarden2021formation}.
The properties of every individual gravitational-wave event
    can be incorporated into a model of the compact binary 
    merger population, 
    which has implications for the astrophysical significance of
    these formation channels
    \cite{LIGO-O2-RP, LIGO-O3-O3a-RP, LIGO-O3-O3b-RP,
    Wysocki2018, Wysocki2019}.

In this dissertation,
    I demonstrate an efficient characterization of the properties
    of the gravitational-wave observations in the Gravitational-Wave
    Transient Catalogs,
    in order to constrain the processes by which 
    massive stellar binaries evolve to become merging 
    compact binaries.
This work builds upon my existing publications
    \cite{nal-chieff-paper, nal-methods-paper}.
We make use of the StarTrack code
    to evaluate constraints on binary evolution
    using simulated populations of merging binaries.

\begin{section}{Characterizing Gravitational-Wave Observations}
\label{sec:gw-properties}

Models describing a gravitational-wave signal are 
    characterized by some parameters, $\BinaryParameters$, consisting of 
    some \textit{extrinsic} parameters representing the sky location,
    distance, and the orientation of the system,
    as well as \textit{intrinsic} parameters which fully characterize
    the astrophysical properties of an event without reference to the observer.
The intrinsic parameters we study in our work
    are the mass ($m_i$, in units of solar mass, $M_{\odot}$), 
    dimensionless spin ($\boldsymbol{\chi_i}$),
    and tidal deformability of neutron stars
    ($\Lambda_1$, $\Lambda_2$).
The \textit{astrophysical} parameters include the intrinsic parameters
    as well as luminosity distance ($\lumdist$, in units of $\mathrm{Mpc}$),
    as the age of the Universe at the time of merger is 
    a quantity of astrophysical interest.

Additional parameterizations are useful for more precisely
    measuring the properties of a gravitational-wave observation,
    due to the sensitivity of the waveform,
    a desire for Gaussian uncertainty,
    and coordinate degeneracies.
These parameters include
    the total mass $M = m_1 + m_2$,
    the symmetric mass ratio $\eta = (m_1 m_2) M^{-2}$,
    the chirp mass $\mathcal{M}_c = \eta^{3/5}M^{2/5}$,
    and inverse luminosity distance $\lumdist^{-1}$.
Except where specified, mass parameters are characterized in the source-frame.
The relationship between source-frame and detector-frame mass parameters
    depends on cosmological redshift: $M_{\redshift} = M_{s} (\redshift + 1)$.
We make use of the standard effective spin parameter which is aligned with
    the orbital angular momentum of the system, $\mathbf{L}$
    \cite{Damour2001, Racine2009, Ajith2011}:
\begin{align}\label{eq:chi_eff}
    \chi_{\textrm{eff}} = 
        (\boldsymbol{\chi}_1m_1 + \boldsymbol{\chi}_2m_2) \cdot
        \hat{\mathbf{L}}/M
\end{align}
We make use of the standard Cartesian coordinate system defined such that
    $\hat{z}$ is orthogonal to the plane of the orbit
    at the reference time (frequency) at which
    orbital initial data is specified
    (i.e. $\hat{z} = \hat{L}$).

The likelihood $\mathcal{L}(h(\BinaryParameters)) \equiv \mathcal{L}(\BinaryParameters)$
    of a known gravitational-wave signal $h(t|\BinaryParameters)$ can be evaluated
    in a parameter space, $\BinaryParameters$,
    based on some knowledge of the detector noise.
Detailed models for gravitational-wave emission provide estimates
    for $h(t|\BinaryParameters)$
    \cite{Pankow2015,Veitch2015LALInference, 
    2019PASP..131b4503B, RIFT, Bilby2019, IAS,
    GW150914-num}.
The Bayesian problem of gravitational-wave parameter inference
    is to characterize the posterior probability 
    of each event as a function of its parameters
    \cite{Bayes}.
This requires a careful and thorough investigation of
    the prior information characterized by a fiducial prior,
    $p(\BinaryParameters)$.
The posterior probability is therefore proportional to
    $\mathcal{L}(\BinaryParameters)p(\BinaryParameters)$.

\correction{
Many groups are working on characterizing each event
    using full numerical relativity
    \cite{HealyNR, GW150914-num, Ramos-Buades-NR,
    NINJA, NRAR, SXS2013, ETK2012, ETK2013,
    Husa-NR, Boyle_2019_SxS, Jani_2016_GA_Tech, BAM2008}.
}
While a full numerical relativity approach is necessary to 
    simulate precessing compact binaries during inspiral
    and merger in complete rigor,
    many groups including the LSC use approximants
    which can be evaluated more quickly on the scale 
    required for parameter estimation.
\correction{
These approximants include 
    IMRPhenomD \cite{IMRPhenomDa, IMRPhenomDb},
IMRPhenomPv2 \cite{IMRPhenomPv2, IMRPhenomPv2b},
    IMRPhenomPv3 \cite{IMRPhenomPv3},
    IMRPhenomPv3HM \cite{IMRPhenomPv3HM},
    IMRPhenomXPHM \cite{IMRPhenomXPHMa, IMRPhenomXPHMb, IMRPhenomXPHMc},
    SEOBNRv3 \cite{SEOBNRv3a, SEOBNRv3b},
    SEOBNRv4 \cite{SEOBNRv4},
    SEOBNRv4PHM \cite{SEOBNRv4PHMa, SEOBNRv4PHMb},
    and NRSur7dq4 \cite{NRSur7dq4},
    as well as others 
    \cite{TEOBResumSa, TEOBResumSa,
    NRTidalExt, NRTidalv2Ext, SEOBNRv4Ta, SEOBNRv4Tb}.
}

One such parameter estimation framework developed by my colleagues at RIT
    is the Rapid parameter inference on gravitational-wave sources
    via Iterative FiTing (RIFT) algorithm,
    which addresses the challenge of iteratively fitting the posterior
    probability for gravitational-wave sources in some set of parameters
    \cite{RIFT}.
This is accomplished by choosing points in that parameter space and 
    evaluating an expensive likelihood function for the waveform
    best characterizing a given signal.
Those likelihood evaluations are then interpolated
    and weighted by the fiducial prior,
    in order to reconstruct the posterior probability function,
    in the form of a large set of samples ($\BinaryParameters_k$)
    representing fair draws which fully characterize that posterior.

Several key science objectives of gravitational-wave astronomy
    explore the astrophysical implications of the entire population
    of observed compact binary mergers.
These \emph{population synthesis} calculations are often
    Monte Carlo simulations of millions of synthetic merging binaries.
To assess the net likelihood of a specific population synthesis model
    therefore often requires large scale Monte Carlo integrals
    of the likelihood of each gravitational-wave event
    \cite{Talbot2019GWPopulation, Wysocki2019,
    LIGO-O3-O3b-RP, Belczynski2020, Breivik_2020,
    Sadiq2021, Edelman2021, Tiwari_2021,
    Veitch2015LALInference, Bilby2019}.
As the sensitivity of the detectors increases towards design sensitivity
    in the coming years, these calculations will become more and more
    computationally expensive as the population of gravitational-wave
    events grows
    \cite{fritschel2020instrument}.
For other applications, an evaluation of $\mathcal{L}(\BinaryParameters)$ is
    required, which sample-based methods may fail to provide.

There are many methods for reconstructing
    this likelihood function, including sample-based
    methods such as Kernel Density Estimates (KDEs)
    and histograms \cite{Ghosh_2021}.
Such sample-based methods are useful for low-dimensional models,
    but often scale poorly to higher-dimensional use cases,
    requiring an increasing number of samples to fully 
    characterize the intrinsic parameters of an event.
Carefully tuned non-parametric methods
    can protect models from binning and smoothing effects
    (for example, see \cite{Golomb_2022}
        with the binary neutron star merger, GW170817).
Compressed parametric models, such as the multivariate normal
    distribution pose a viable alternative to sample-based methods
    \cite{cho2013,GW150914-num,RIFT,jaranowski2007gravitationalwave}.
Fast and accurate approximations for
    the likelihood function $\mathcal{L}(\BinaryParameters)$ are essential
    to carrying out the central calculation of 
    a population synthesis algorithm.

The multivariate normal distribution has long been used to describe
    the likelihood of gravitational-wave events in coordinates
    well suited to waveform models.
However, a truncated set of samples can introduce bias
    in a sample-based parameterization of the multivariate
    normal distribution \cite{nal-chieff-paper, nal-methods-paper}.
In our previous work, we have introduced bounded multivariate Normal
    Approximate Likelihood (NAL) models for each event,
    which fully characterize the precessing degrees of freedom
    of each gravitational-wave event in the GWTCs
    \cite{nal-chieff-paper, nal-methods-paper}.
These models overcome bias introduced by finite boundary effects
    and are efficient in both generating samples and evaluating
    a likelihood estimate.
They lend themselves immediately to both population work and
    low-latency gravitational-wave parameter estimation.

\end{section}

\begin{section}{Origins of Compact Binary Objects}

Theories of how compact binary objects form include
    dynamic mergers in a dense stellar cluster,
    as well as 
    the binary evolution of massive stars,
    with some mechanism for ejecting orbital angular momentum from the system
    \correction{
    \cite{Belczynski2016Nature,Silsbee2017,Gayathri2022,Bartos_2017,
    Miller_2002,Portegies_Zwart_2002,Portegies_Zwart_2004,OLeary2009,
    Kocsis2012, Naoz_2013, Antognini2014a, Antonini_2014b, Morscher_2015}.}
With some noteworthy exceptions, such as the BNS observation,
    GW170817, gravitational-wave detections often are not accompanied
    by an electromagnetic counterpart
    \cite{abbott2017gw170817, coulter2017swope, cowperthwaite2017electromagnetic,
        kilpatrick2017electromagnetic, shappee2017early, kasen2017origin}.
Most often, we must infer the astrophysical origin of 
    compact binary sources from the gravitational-wave signal alone.
Many studies have shown that isolated binary evolution can produce most of the
    events published thus far in the GWTCs
    \cite{1991ApJNarayan, Belczynski2002, Belczynski2008, Belczynski2016,
        Belczynski2020, broekgaarden2021formation, posydon}.
Exceptions to this would include events with a high orbital eccentricity
    and precessing spin components, such as proposed for events
    like GW190521 in the third LIGO observing run
    \cite{Gayathri2022, LIGO-O3-O3a-RP, LIGO-O3-O3b-RP}.
In the remainder of this work, we focus on the isolated binary evolution
    formation channel for compact binary objects.

As most compact binary objects are expected to originate from isolated
    binary evolution, the population of gravitational-wave signals 
    detected so far allow us to constrain the bulk properties of 
    this binary evolution formation channel.
This is accomplished by comparing the real gravitational-wave
    observations from published catalogs to predicted
    populations of compact binary mergers, with
    different assumptions about binary evolution
    using a Bayesian inference framework
    \cite{Bayes}.
In our work, we use the NAL models for the likelihood
    of each event in the Gravitational-Wave Transient Catalogs
    to carry out this inference.
We incorporate these models together with the
    StarTrack population synthesis code for the 
    prediction of compact binary mergers from
    isolated binary evolution and an associated
    cosmological postprocessing setup
    \cite{DominikI, DominikII, DominikIII, Belczynski2020}.
\end{section}

\begin{section}{Organization of this Dissertation}

In chapter \ref{chap:nal} I describe the Normal Apprixmate Likelihood (NAL)
    models for the likelihood of gravitational-wave events introduced
    in Delfavero et al. (2021 and 2022) 
    \cite{nal-chieff-paper,nal-methods-paper}.
Chapter \ref{chap:binary} reviews some of the pertinent background
    and literature for
    the isolated binary evolution models we constrain
    as well as the formulation for predicting the gravitational-wave
    detection rate for simulated synthetic universes with
    the StarTrack code.
Following this, chapter \ref{chap:popsyn} demonstrates the
    Bayesian framework used to constrain these processes
    by comparing individual simulations with different
    sets of formation parameters.
Finally, chapter \ref{chap:conclude} discusses
    the astrophysical implications of our research.

This dissertation summarizes the work in two of my previous papers,
    including Delfavero et al. (2021) \cite{nal-chieff-paper} and
    Delfavero et al. (2022) \cite{nal-methods-paper}.
I have made contributions to RIFT in the past,
    such as in my master's thesis \cite{DelfaveroMastersThesis}.
While I continue to contribute to parameter estimation efforts,
    such as in the forthcoming paper by Wofford et al. (2022)
    \cite{RIFTUpdate},
    these are not the focus of this dissertation.
My contributions to efforts to constrain the neutron star
    equation of state are also not the focus of this dissertation
    \cite{holmbeck2021nuclear},
    although this work will be discussed briefly as an example of the
    astrophysical consequences of our population synthesis.

\end{section}

 \end{chapter}

\begin{chapter}{Normal Approximate Likelihoods for Gravitational Wave Events}
\label{chap:nal}

As described in section \ref{sec:gw-properties},
    each gravitational-wave signal detected by an observatory connected to LVK
    is observed as a strain incident on the detector
    \cite{O3-Detector, GW150914-detection}.
In order to characterize the properties of each event,
    parameter estimation groups use Bayesian inference,
    and compare this strain information to relativistic waveform models
    described by a set of parameters (including the astrophysical
    parameters $\BinaryParameters$)
    \cite{Pankow2015,Veitch2015LALInference, RIFT, Bilby2019, IAS, Bayes}.
In publications such as the Gravitational-Wave Transient Catalogs,
    these parameter estimates are released as a set of identically distributed
    independent random samples from the posterior of each event,
    as well as single-valued parameter estimates which provide
    single-parameter credible intervals (i.e., from their one-dimensional
    marginal distributions)
    \cite{GWTC-1, GWTC-2, GWTC-2p1, GWTC-3,
    IAS, 3-OGC, Boyle_2019_SxS, Jani_2016_GA_Tech, BAM2008,
    GWTC-2p1-Zenodo, GWTC-3-Zenodo}.
In this work we focus on GWTC events, however these methods can
    be applied to any sample-based estimate for the parameters
    of an arbitrary probability distribution.

In low dimensionality
    sample-based estimates of the likelihood function,
    such as a Kernel Density Estimate (KDE),
    can provide an accurate representation
    of the likelihood function for a gravitational-wave event
    \cite{nal-chieff-paper, Ghosh_2021}.
These sample-based estimates reconstruct the likelihood by re-weighing
    posterior samples by the inverse prior 
    (see section \ref{sec:sample-density}).
These methods are ultimately limited in higher-dimensional parameter spaces
    by the number of samples required to construct an accurate approximation,
    which increases with dimensionality.
Therefore, as sample-based methods such as this are restricted
    to the parameter estimation samples released with each catalog,
    and the computational cost of these sample-based estimates
    increases with the number of samples,
    these methods are not suitable to large-scale population inference.

\begin{figure}
\centering
\includegraphics[width=0.49\textwidth]{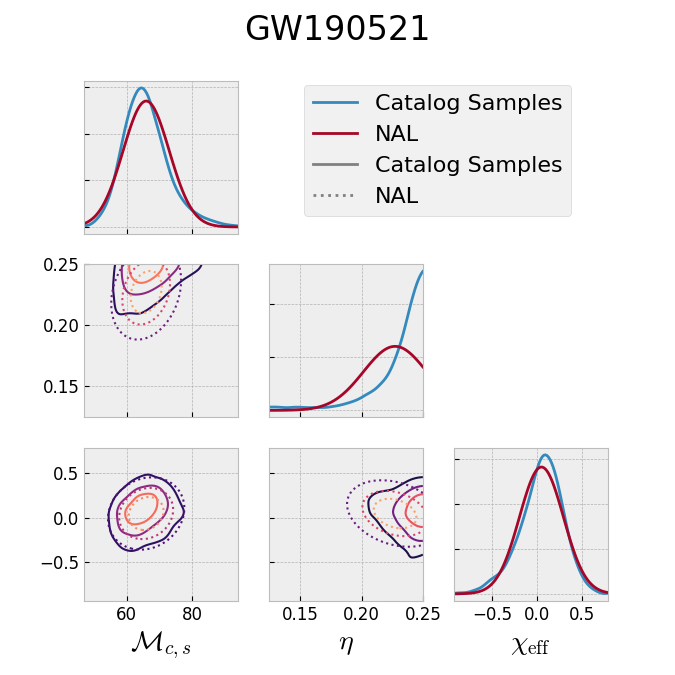}
\includegraphics[width=0.49\textwidth]{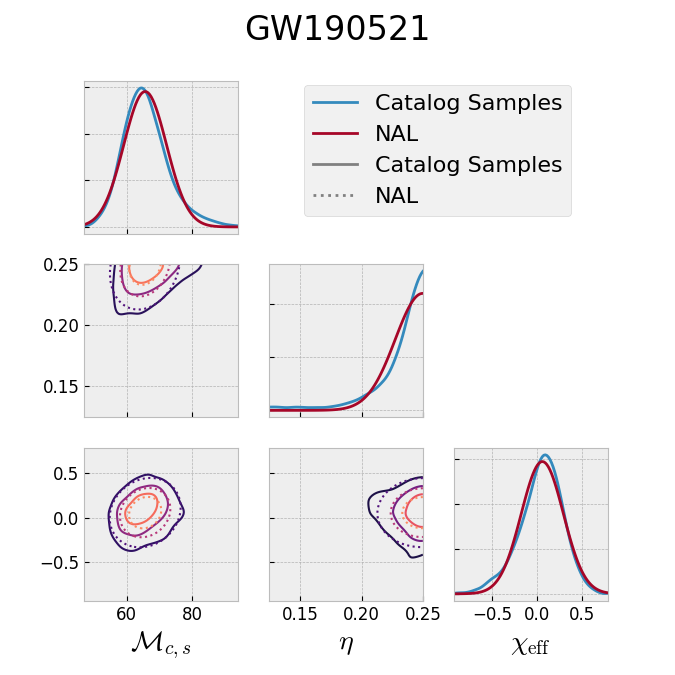}
\caption{\label{fig:nal-justification}
NAL models for GW190521 \newline
    (https://dcc.ligo.org/public/0169/P2000223/007/GW190521.tar
    - using the `PublicationSamples' set of samples),
    using \textit{implicit} (left) and
    \textit{optimized} (right) fit parameters.
These corner plots evaluate the inferred likelihood function in one 
    (diagonal) and two (off-diagonal) sample spaces.
In the one-dimensional spaces, the blue curve indicates
    a non-parametric estimate of published samples,
    while the red curve indicates the properly re-normalized Gaussian.
In the two-dimensional sample spaces, contours are drawn enclosing
    $[25, 50, 75]$ percent of the likelihood.
Note that optimized NAL models overcome the boundary effect
    present for the implicit fit near equal mass ($\eta = 0.25$).
\correction{
The one- and two-dimensional sample estimates which are compared to each
    Gaussian are the marginal density estimates described in section
    \ref{sec:gp-api}
}}
\end{figure}

As adopting a likelihood function from these samples can be non-trivial,
    and as the sample mean often fails to describe the
    Maximum Likelihood Estimate (MLE) for the parameters of each gravitational
    wave event,
    we explore alternative representations of the same parameter estimation
    samples through bounded (truncated) Normal Approximate Likelihood (NAL) 
    models
    \cite{nal-chieff-paper, nal-methods-paper}.
These models overcome limitations imposed by a finite set of samples,
    to provide computationally efficient likelihood approximations
    which are not restricted by the same finite boundary effects as 
    non-truncated Gaussians
    (consider samples of the symmetric mass ratio for an equal mass event,
    e.g. fig \ref{fig:nal-justification}).
The parameters of each NAL model also describe a less biased MLE
    description of the astrophysical properties of each event.

In this chapter, I will describe my methods for using data products 
    such as the GWTC parameter estimation samples
to approximate the likelihood function for each published
    gravitational-wave event in its full set of astrophysical parameters.
I will also describe the immediate astrophysical consequences
    following the NAL models,
    such as the properties of our gravitational-wave population.
Finally, I will discuss several potential applications of these methods.
The work in this chapter relies heavily on the associated
    papers \cite{nal-chieff-paper, nal-methods-paper}.

\begin{section}{Constructing the Likelihood from PE Samples}
\label{sec:sample-density}

When using samples to construct a general parametric model for some density,
    it is useful to first construct a non-parametric model
    which is accurate
    (even if comparatavely more expensive to evaluate).
Following this,
    one can generate a set of evaluations to compare with
    the parametric model.
This section highlights the non-parametric Gaussian Process (GP) models
    for the one- and two-dimensional marginalization of each
    set of parameters.
These GP models are fully generic,
    not assuming Gaussian behavior,
    providing efficient and accurate approximations
    to the sample density,
    and which may themselves be desired for additional applications.

Reconstructing a likelihood density from the posterior samples
    first requires re-weighting samples by the inverse prior
    \cite{Bayes, RIFT}.
The GWTC releases use an uninformed prior
    which is uniform in detector-frame component masses
    and spin magnitude and direction
    \cite{CallisterPrior2021}.
Alternatively, in terms of 
    $\mathcal{M}_{c,\redshift}$ (detector-frame chirp mass) and $\eta$:
\begin{equation}\label{eq:mass-prior}
p(\mathcal{M}_{c,\redshift}, \eta) \mathrm{d}\mathcal{M}_{c,\redshift} \mathrm{d}\eta =
    \frac{4}{(M_{\mathrm{max}} - m_{\mathrm{min}})^2}
    \frac{\mathcal{M}_{c,\redshift}\mathrm{d}\mathcal{M}_{c,\redshift}\mathrm{d}\eta}{
        \eta^{6/5}\sqrt{1 - 4 \eta}}
\end{equation}
where we have for simplicity adopted prior boundaries
    in $\mathcal{M}_{c,\redshift}$ and $\eta$ so the distribution takes a simple
    product form.
When applicable, the distance prior we remove is
    the square of the inverse luminosity distance
    $1/\lumdist^2$.
Our method is not limited to an uninformed prior,
    and the same methods could be applied for posterior
    samples drawn with an informed prior
    \cite{Vitale:2017cfs}.

There are various methods for providing a sample-based
    estimate of the likelihood function.
In low dimensionality, a KDE provides an accurate estimate
    of the likelihood function for a gravitational-wave,
    and we use KDEs in our first paper, as the intermediate
    approximation to the likelihood
    \cite{nal-chieff-paper, Ghosh_2021}.
However, the evaluation of a KDE for a given point in parameter space
    requires a high computational cost, as the evaluation
    scales with the number of samples used for the estimate.
Some events in LIGO's third observing run have $> 10^5$ samples,
    which make KDEs less than ideal \cite{GWTC-2}.

Histograms provide a simple sample-based method
    but require smoothing and careful handling to avoid overfitting 
    or underfitting samples.
Others have done a lot of work developing adaptive binning
    algorithms for flexible histograms which try to avoid
    these errors by using different bin widths throughout a sample
    \cite{ScargleHistogram2012}.
Some of these methods are used in gravitational-wave
    likelihood estimation
    \cite{Cornish2010, Zackay2018, Cornish2021, Leslie2021}.
These methods are well tested in one dimension, are computationally
    limited in higher dimensions.
Some of these methods are still prone to overfitting.

In order to recover the marginal likelihood for our parameters
    in one and two dimensions,
    we rely on histograms with a fixed bin width
    optimized by minimizing binning effects.
We smooth these histograms using a Gaussian Process,
    and we share our methods below:

\begin{subsection}{Gaussian Process Marginal Densities}
\label{sec:gp-api}

\correction{
Gaussian Process Regression is a non-parametric method of approximating
    the value of a function in a given parameter space,
    given some sample function values (``training data'')
    \cite{williams2006gaussian}.
GP methods use a kernel, or a set of basis functions which
    describe the covariance of points in the space explored by the model.
The function value at 
    each point in the space described by a Gaussian Process
    can be described by a multivariate normal random variable.
These methods can be reliable, and 
    are useful in statistical inference due to their ability
    to make predictions directly by providing
    a mean and covariance for the estimated value of a given function.

GP methods are known to be highly accurate while also being
    computationally expensive,
    where the time required to evaluate the function on a sample
    scales with the cube of the number of training points provided
    in traditional methods.
This cost can be mitigated
    by selecting a kernel that enables sparse matrix operations,
    allowing the evaluation of the GP on a given sample
    to consider only the points closest to that sample.
Williams and Rasmussen provide such a kernel in the form of
    the piecewise polynomial kernel with compact support
    \cite{williams2006gaussian}.
Together with compiled evaluations of the basis functions themselves,
    the computational cost can be reduced by several orders of magnitude
    compared to sci-kit learn \cite{scikit-learn}.
Our Gaussian Process code is available at \\
    \url{https://gitlab.com/xevra/gaussian-process-api}.
This package performs a Gaussian Process fit
    in an arbitrary number of dimensions
    using compiled sparse-matrix kernel operations
    and sparse Cholesky matrix inversion.
Our package also provides a whitenoise kernel,
    necessary for many applications.

In addition to the Gaussian Process Regression interpolator,
    the package provides a way to construct an estimate to 
    a weighted sample density function,
    as a set of one- and two-dimensional marginal density estimates,
    or as an k-dimensional density estimate.
The k-dimensional estimate is not required for our applications,
    and requires additional testing,
    so we leave it out of this discussion.
}

Our method for reconstructing the one- and two- dimensional 
    likelihood from re-weighted posterior samples
    begins with a histogram with some number of bins, $n$.
We consider the centers of each bin our training inputs, $X_n$,
    and our normalized histogram values make up our training values
    $Y_n$.
A Gaussian Process is used to smooth histogram values,
    and is labeled $f_n(\BinaryParameters)$,
    using standard binomial error estimates.
We then construct a similar histogram and Gaussian Process
    $f_{n+1}(\BinaryParameters)$, using $n + 1$ bins.
\correction{
For two- and higher-dimensional histograms,
    all bins are incremented together - but need not be identical.
}
We cross-evaluate our models to furhter 
    estimate the error of each model,
    which is summed in quadrature with binomial error estimates.

\begin{align}
\delta Y_{n} = \abs{Y_{n} - f_{n+1}(X_n)} \\
\delta Y_{n + 1} = \abs{Y_{n+1} - f_{n}(X_{n+1})}
\end{align}

By incrementing the number of bins, 
    we choose $n$ to minimizes these error estimates,
    using a Kolmogorov-Smirnov test \cite{kolmogorov1933sulla}.
\correction{
The Kolmogorov-Smirnov test measures the maximum value
    of the difference between two distributions.
We refrain from using a more sophisticated measure of the difference
    between two distributions in order to avoid over-complicating
    a new method.
In the future, a different test may be used.
}

\correction{
One limitation of this method without any additional work
    is that the training data ($X_n$)
    are placed at the centers of histogram bins.
Therefore, there will be no training points in the outermost
    half-bin-width region of the histogram.
If we use $f_{n}(X)$ to estimate the value of the likelihood function
    in that outermost half-bin-width region, we 
    are required to extrapolate rather than interpolate.
This is more difficult to accomplish accurately with
    a Gaussian Process not designed specifically for
    extrapolation.

We can overcome this limitation by imposing a boundary condition
    on our histogram, taking advantage of the fact that histograms
    estimate a probability distribution only in a bounded region.
This boundary condition effectively mirrors the sample about the boundary in
    only a half-bin-width region outside the histogram limits.
Extending the influence of the histogram by one half-bin-width
    allows the outer-most bins for our histogram to lie precisely
    on the boundary.
This allows us to avoid extrapolating with our Gaussian Process.
}
Finally, we construct an additional Gaussian Process,
    $g(\BinaryParameters)$, which is trained using the combined training sets
    and estimated errors for $f_n(\BinaryParameters)$ and $f_{n+1}(\BinaryParameters)$,
    with our final choice of $n$.
We find that $g(\BinaryParameters)$ offers a more informed Gaussian Process than
    those generated by individual histograms.

\end{subsection}
\end{section}

\begin{section}{Normal Approximate Likelihood Modeling} 
\label{sec:nal-model}

The multivariate normal distribution is a standard tool for
    effective approximations for the gravitational-wave likelihood
    function in suitable coordinates
    \cite{GW150914-num, PoissonGW1995, RichardPEFisher2014, 
        ChoFisher2013, RichardPEPrecessing2014}.
Multivariate normal models have many desirable properties,
    including fast density evaluation
    and ease of generating random samples.
However, a normal approximation does not arise naturally in
    generic coordinates, but only when using a parameterization
    well suited to the signal, such that the coordinates chosen
    are expected to be dominated by Gaussian noise.
Unfortunately, many of these desired coordinates are limited by
    a finite domain (see fig \ref{fig:nal-justification}).
In cases where the likelihood appears Gaussian-like
    with significant truncation,
    an optimized bounded multivariate normal distribution will 
    preserve the desired properties of a Gaussian model
    and better represent a sample density than
    simply assuming the sample mean and covariance for the parameters
    of a Gaussian.
This section summarizes thee parameters and optimization process for the
    Normal Approximate Likelihood (NAL) model.

\begin{subsection}{The NAL Model Parameters}
\label{sec:nal-parameter}

A generalized multivariate normal distribution can be characterized fully
    by the location of its peak ($\mathbf{\mu}$) and a covariance
    matrix ($\mathbf{\Sigma}$).
\begin{align}\label{eq:multivariate-normal}
G(\BinaryParameters-\mathbf{\mu}, \mathbf{\Sigma}) = 
    (|2\pi \mathbf{\Sigma}|)^{-\frac{1}{2}}
    \exp \Big[
    -\frac{1}{2} (\BinaryParameters - \mathbf{\mu})^T \mathbf{\Sigma}^{-1}(\BinaryParameters-\mathbf{\mu})
    \Big]
\end{align}

The NAL parameterization of these quantities includes a decomposition
    of $\mathbf{\Sigma}$ into
the characteristic standard deviation parameters, $\mathbf{\sigma}$,
    and the correlation matrix, $\mathbf{\rho}$,
    where
    $\mathbf{\Sigma} = \mathbf{\sigma} \mathbf{\rho} \mathbf{\sigma}^T$.
This decomposition is especially effective because
    the correlation parameters have useful properties,
    including symmetry
($\rho_{i,j} = \rho_{j,i}$), 
    unity along the diagonal ($\rho_{i,i} = 1$),
    and bounds ($-1 \le \rho_{i,j} \le 1$).
The expression for a $k$-dimensional space expands to:

{
\small
\begin{equation}\label{eq:cov_expansion}
\begin{pmatrix}
    \sigma_1^2 && \sigma_1\sigma_2 \rho_{1,2} && \cdots && \sigma_1\sigma_k\rho_{1,k} \\
    \sigma_1\sigma_2\rho_{1,2} && \sigma_2^2  && \cdots && \sigma_2\sigma_k\rho_{2,k} \\
    \vdots && \vdots && \ddots && \vdots \\
    \sigma_1\sigma_k\rho_{1,k} && \sigma_2\sigma_k\rho_{2,k}  && \cdots && \sigma_k^2 \\
\end{pmatrix}
=
\begin{pmatrix}
    \sigma_1 \\
    \sigma_2 \\
    \vdots \\
    \sigma_k \\
\end{pmatrix}
\begin{pmatrix}
        1 && \rho_{1,2} && \cdots && \rho_{1,k} \\
        \rho_{1,2} && 1 && \cdots && \rho_{2,k} \\
        \vdots && \vdots && \ddots && \vdots \\
        \rho_{1,k} && \rho_{2,k} && \cdots && 1 \\
\end{pmatrix}
\begin{pmatrix}
    \sigma_1 \\
    \sigma_2 \\
    \vdots \\
    \sigma_k \\
\end{pmatrix}^{\mathrm{T}}
\end{equation}
}
where, $(k^2 + 3k)/2$ parameters must be optimized
    (e.g. a three-dimensional model will have nine parameters,
    while an eleven-dimensional model will have seventy-seven parameters).

\end{subsection}
\begin{subsection}{NAL Optimization}\label{sec:nal-optimization}

As one- and two-dimensional marginals fully capture the behavior of
    a generalized bounded multivariate normal distribution,
    we can optimize a high-dimensional NAL model
    by evaluating the goodness of fit
    using one- and two-dimensional marginalization.
A one-dimensional marginal in the i'th dimension is fully characterized by
    $\mu_i$ and $\sigma_i$, as $G_{i}(\BinaryParameter_i) =G(\BinaryParameter_i - \mu_i, \sigma_i)$.
Similarly, a two-dimensional marginal in the i'th and j'th 
    components is fully characterized by $\mu_i$, $\mu_j$, $\sigma_i$, $\sigma_j$, 
    and $\rho_{ij}$, as 
    $G_{i,j}(\BinaryParameter_{i, j}) = G([X_i, X_j] - [\mu_i, \mu_j],
        [[\sigma_i^2,\sigma_i\sigma_j\rho_{ij}],
        [\sigma_i\sigma_j\rho_{ij},\sigma_j^2]])$.
By evaluating $G_{i}(\BinaryParameter_{i})$ and $G_{i,j}(\BinaryParameter_{i,j})$ in a bounded region,
    and normalizing in that bounded region,
    one can invoke a standard convergence test
    to evaluate the goodness of fit between a set of Gaussian parameters
    and an estimate of the sample density.

In our procedure,
    we use the \textbf{Kullback-Leibler (KL) divergence} to evaluate
    the agreement of our NAL models with the 
    Gaussian Process sample density estimates outlined previously
(sec \ref{sec:gp-api}).
\correction{
The KL divergence measures the ability of a secondary probability
    distribution ($Q(x)$) to describe a primary probability distribution
    ($P(x)$), and is given by: 
\begin{align}\label{eq:kl-div}
D_{\mathrm{KL}}(P|Q) = \sum\limits_{x \in \mathbf{x}} P(x) \mathrm{log}
\Big(\frac{P(x)}{Q(x)}\Big)
\end{align}
It also represents the information lost by approximating
    $P(x)$ with $Q(x)$.
The KL divergence is sensitive to logarithmic differences between $P$ and $Q$.
The quantity is well suited to our purposes,
    as we compare a wide variety of guesses (as secondary distributions)
    to fixed (primary) one and two-dimensional marginal sample distributions
    during optimization.
The quality of a particular guess in the optimization
    is described by the mean of the one- and two-dimensional KL divergences
    which describe the one- and two-dimensional marginals of
    the NAL model.
}

\correction{
Initially, we explored various packaged optimization routines,
    such as Newton's method and Emcee \cite{emcee}.
We found that these optimization routines failed to 
    perform at a level we were satisfied with.
Following this, we explored simulated annealing as an alternative
    optimization algorithm, which was used in our first 
    paper \cite{simulated-annealing, nal-chieff-paper}.
We have continued to refine the optimization algorithm
    to include informed guessing and develop a genetic optimization routine.
These ``informed guesses'' include quadratic fits to one-dimensional
    marginals, for estimating $\mu$ and $\sigma$ in each dimension.

While the final model is selected using the mean of the
    one- and two-dimensional KL divergences,
    we are able to make use of the individual KL divergences
    for each marginal.
The initial population of the genetic algorithm is a set of informed guesses,
    filled with additional random guesses.
At each time-step,
    \textit{parents} are selected by considering the mean KL divergence.
The \textit{traits} passed down to the child population are the
    parameters of the multivariate normal fit,
    and are chosen randomly from the two parents
    \textit{with a preference for the lower one- and two-dimensional
    KL divergences of associated marginals}.
Following this,
    the child population is allowed to jump like a 
    Metropolis-Hastings Markov Chain Monte Carlo (MCMC) random walker,
    which constitutes \textit{mutation}
    \cite{MetropolisHastings}.
At each step,
    the guesses with the lowest mean KL divergence are carried over
    to the breeding pool of the next generation.
}

Alongside our second paper, we provide a software package, GWALK
    (Gravitational-Wave Approximate LiKelihoods),
    at \url{https://gitlab.com/xevra/gwalk}.
This package provides tools for constructing NAL
    for an arbitrary set of samples,
    within an arbitrary set of boundaries.
It also provides a computationally efficient
    compiled Gaussian likelihood evaluation
    and our genetic optimization routine
    for NAL parameters.

\end{subsection}
\end{section}

\begin{section}{Properties of NAL Models for GWTC Events}

We find that NAL models for most events in the GWTCs
    are overwhelmingly in agreement with
    their associated sample density.
These optimized models better represent the likelihood inferred
    from parameter estimation samples than
    Gaussians with assumed parameters based on the 
    sample mean and covariance
    while providing the same efficient density evaluation
    and ease in generating randomly distributed samples.
Additionally,
    the $\mathbf{\mu}$ parameters of these models provide a more accurate
    MLE description of the astrophysical parameters for a given event
    than either the sample mean or median when significant boundary/truncation
    effects are in play.
This is immediately relevant to astrophysical interpretations
    of gravitational-wave events,
    as boundary effects extend beyond a single parameter,
    and are unavoidable when fully characterizing the astrophysical
    parameters of a gravitational-wave event.
The NAL model fits for the full parameterization
    of each waveform in the GWTC releases
    are available at \url{https://gitlab.com/xevra/nal-data}.
The rest of this chapter considers the application and interpretation of
    these NAL models.

\begin{table}
\centering
{
\small
\begin{tabular}{|l|l|l|}
\hline
Parameterization & Coordinates & Prior \\
\hline
aligned3d                   & $\mathcal{M}_{c,z}$, $\eta$, $\chi_{\mathrm{eff}}$ & aligned3d \\
aligned3d\_source           & $\mathcal{M}_{c}$, $\eta$, $\chi_{\mathrm{eff}}$ & aligned3d \\
aligned3d\_dist             & $\mathcal{M}_{c,z}$, $\eta$, $\chi_{\mathrm{eff}}$, $\lumdist^{-1}$ & aligned3d\_dist \\
mass\_tides                 & $\mathcal{M}_{c,z}$, $\eta$, $\tilde{\Lambda}$, $\delta \tilde{\Lambda}$ & mass\\
mass\_tides\_source         & $\mathcal{M}_{c}$, $\eta$, $\tilde{\Lambda}$, $\delta \tilde{\Lambda}$ & mass \\
aligned\_tides              & $\mathcal{M}_{c,z}$, $\eta$, $\chi_{\mathrm{eff}}$, $\tilde{\Lambda}$, $\delta \tilde{\Lambda}$ & aligned3d \\
aligned\_tides\_source      & $\mathcal{M}_{c}$, $\eta$, $\chi_{\mathrm{eff}}$, $\tilde{\Lambda}$, $\delta \tilde{\Lambda}$ & aligned3d \\
aligned\_tides\_dist        & $\mathcal{M}_{c,z}$, $\eta$, $\chi_{\mathrm{eff}}$, $\tilde{\Lambda}$, $\delta \tilde{\Lambda}$, $\lumdist^{-1}$ & aligned3d\_dist \\
spin6d                      & $\chi_{1x}$, $\chi_{2x}$, $\chi_{1y}$, $\chi_{2y}$, $\chi_{1z}$, $\chi_{2z}$ & precessing8d \\
precessing8d                & $\mathcal{M}_{c,z}$, $\eta$, $\chi_{1x}$, $\chi_{2x}$, $\chi_{1y}$, $\chi_{2y}$, $\chi_{1z}$, $\chi_{2z}$ & precessing8d \\
precessing8d\_source        & $\mathcal{M}_{c}$, $\eta$, $\chi_{1x}$, $\chi_{2x}$, $\chi_{1y}$, $\chi_{2y}$, $\chi_{1z}$, $\chi_{2z}$ & precessing8d \\
precessing8d\_dist          & $\mathcal{M}_{c,z}$, $\eta$, $\chi_{1x}$, $\chi_{2x}$, $\chi_{1y}$, $\chi_{2y}$, $\chi_{1z}$, $\chi_{2z}$, $\lumdist^{-1}$ & precessing8d\_dist \\
precessing\_tides\_source   & $\mathcal{M}_{c}$, $\eta$, $\chi_{1x}$, $\chi_{2x}$, $\chi_{1y}$, $\chi_{2y}$, $\chi_{1z}$, $\chi_{2z}$, $\tilde{\Lambda}$, $\delta \tilde{\Lambda}$ & precessing8d \\
full\_precessing\_tides     & $\mathcal{M}_{c,z}$, $\eta$, $\chi_{1x}$, $\chi_{2x}$, $\chi_{1y}$, $\chi_{2y}$, $\chi_{1z}$, $\chi_{2z}$, $\tilde{\Lambda}$, $\delta \tilde{\Lambda}$, $\lumdist^{-1}$ & precessing8d\_dist \\
\hline
\end{tabular}
}
\caption{\label{tab:coord_tags}
NAL parameterizations of the astrophysical properties
    associated with parameter estimation samples in the GWTCs.
}
\end{table}

Most of the compact binaries reported in the GWTC releases
    are well constrained by source-frame chirp mass ($\mathcal{M}_c$),
    symmetric mass ratio ($\eta$), and 
    aligned effective spin ($\chi_{\mathrm{eff}}$)
    \cite{LIGO-O2-RP, LIGO-O3-O3a-RP, LIGO-O3-O3b-RP}.
Other parameterizations are necessary to fully characterize events
    with properties different from the typical binary black hole merger,
    such as precessing spin and tidal deformability.
Our choices for the parameterizations used for NAL models of a given event
    are given by table \ref{tab:coord_tags}.

These parameterizations allow us to provide detailed
    and accurate models for the entire astrophysical parameter
    space for the gravitational-wave events reported in the 
    GWTC releases for the third observing run of LIGO/Virgo,
    as well as a limited set of models for events in GWTC-1
    which don't offer full spin information.
    
\correction{
The labels representing the prior removed for each parameterization
    described in table \ref{tab:coord_tags}
    are all consistent with the prior choices described in section
    \ref{sec:sample-density}.
However, the prior removed must represent the coordinates of
    a given parameterization,
    which differ from the aligned spin case to the precessing case,
    and for parameterizations that consider distance.
The `mass' prior removes only the prior in mass.
The `aligned3d' prior removes the prior in mass and $\chi_{\mathrm{eff}}$.
The `precessing8d' prior removes the prior in mass and Cartesian spin.
The `\_dist' suffix indicates that distance prior has been removed,
    as necessary for parameterizations which consider the inverse luminosity
    distance.
}

We have applied the NAL method to each set of samples
    available in the GWTC releases
\correction{
    (GWTC-1: \url{https://dcc.ligo.org/LIGO-P1800370/public},
    GWTC-2: \url{https://dcc.ligo.org/LIGO-P2000223/public/},
    GWTC-2.1: \url{https://zenodo.org/record/5117762#.YwYrG9LMJB1},
    GWTC-3: \url{https://zenodo.org/record/5546665#.YwYrHdLMJB1})
}
    for each of the waveform models available for each set of samples
\correction{
    (including 
    IMRPhenomD \cite{IMRPhenomDa, IMRPhenomDb},
IMRPhenomPv2 \cite{IMRPhenomPv2, IMRPhenomPv2b},
    IMRPhenomPv3 \cite{IMRPhenomPv3},
    IMRPhenomPv3HM \cite{IMRPhenomPv3HM},
    IMRPhenomXPHM \cite{IMRPhenomXPHMa, IMRPhenomXPHMb, IMRPhenomXPHMc},
    SEOBNRv3 \cite{SEOBNRv3a, SEOBNRv3b},
    SEOBNRv4 \cite{SEOBNRv4},
    SEOBNRv4PHM \cite{SEOBNRv4PHMa, SEOBNRv4PHMb},
    and NRSur7dq4 \cite{NRSur7dq4},
    as well as others 
    \cite{TEOBResumSa, TEOBResumSa,
    NRTidalExt, NRTidalv2Ext, SEOBNRv4Ta, SEOBNRv4Tb}
).}
These models support a wide variety of potential
    and realized applications.

\begin{subsection}{Aligned Spin}
\begin{figure}
\centering
\includegraphics[width=0.48\columnwidth]{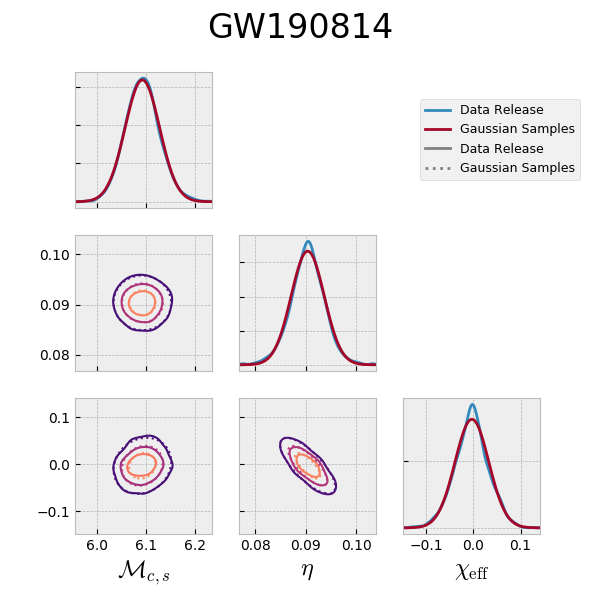}
\includegraphics[width=0.48\columnwidth]{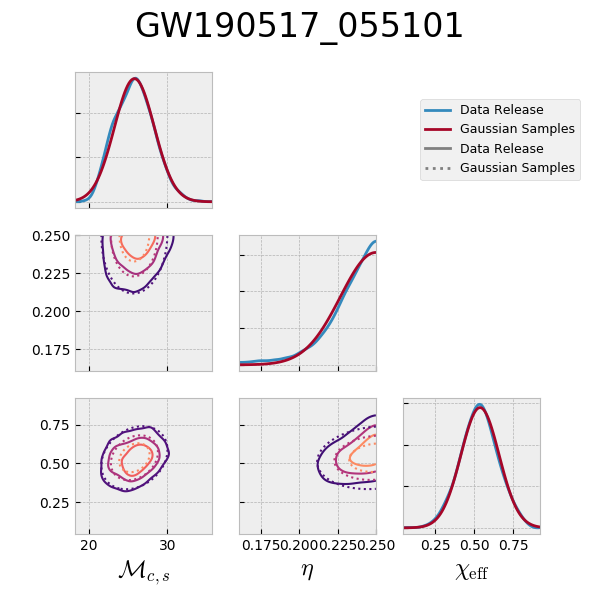}
\caption{\label{fig:nal-aligned3d_source}
These corner plots are similar to figure \ref{fig:nal-justification},
    and represent optimized NAL fits for GW190814 and GW190517\_055101
    (again, through `PublicationSamples').
These are noteworthy events in GWTC-2, with GW190814 having
    a mass ratio significantly different from unity,
    and GW190517\_055101 having a large spin magnitude.
}
\end{figure}

\begin{table}
\centering
{
\small
\begin{tabular}{|l|c|c|c|c|}
\hline
Event & KL & $\mu_{\mathcal{M}_c}$ & $\mu_{\eta}$ & $\mu_{\chi_{\mathrm{eff}}}$ \\
\hline
GW150914 & $0.01$ &$28.6^{+1.1}_{-1.1}$ & $2.500_{-0.007}$ & $0.00^{+0.09}_{-0.09}$ \\
GW170817 & $0.02$ & $1.187^{+0.002}_{-0.002}$ & $0.247^{+0.003}_{-0.003}$ & $0.007^{+0.008}_{-0.008}$ \\
GW190403\_051519 &$0.01$ & $35.5^{+4.9}_{-4.9}$ & $0.17^{+0.03}_{-0.03}$ & $0.76^{+0.09}_{-0.09}$ \\
GW190425 &$0.02$ & $1.43^{+0.01}_{-0.01}$ & $0.228^{+0.017}_{-0.017}$ & $0.09^{+0.05}_{-0.05}$\\
GW190517\_055101 &$0.005$ & $25.8^{+2.5}_{-2.5}$ & $2.50_{-0.02}$ & $0.54^{+0.12}_{-0.12}$ \\
GW190521 & $0.016$ & $65.6^{+6.5}_{-6.5}$ & $2.498^{+0.002}_{-0.022}$ & $0.06^{+0.23}_{-0.23}$ \\
GW190814 & $0.002$ & $6.09^{+0.04}_{-0.04}$ & $0.090^{+0.003}_{-0.003}$ & $0.00^{+0.04}_{-0.04}$ \\
\hline
\end{tabular}
}
\caption{\label{tab:aligned3d_source}
Optimized MLE properties of gravitaional-wave events
    in the aligned-spin source-frame model.
\correction{
The goodness of fit is measured by the mean of the one- and two-dimensional
    KL divergences comparing each NAL model to each marginal
    (see section \ref{sec:nal-optimization}).
}
The SEOBNRv3 approximant is used for GW150914, while the 
    IMRPhenomPv2NRT\_lowSpin approximant is used for GW170817,
    SEOBNRv4PHM is used for GW190403\_051519,
    and ``PublicationSamples'' is used for the remaining events.
}
\end{table}

As seen in figure \ref{fig:nal-aligned3d_source},
    NAL models are in strong agreement with the GWTC samples.
As the majority of events are near equal mass,
    the finite boundary effect of sample truncation 
    in symmetric mass ratio ($\eta \le 1/4$)
    introduces a significant bias away from equal mass in
    Gaussians assumed from the sample mean and covariance
    (fig \ref{fig:nal-justification}).
NAL models overcome this source of bias.
For example, for the gravitational-wave event GW190521,
    we estimate 
    $(\mu_{\mathrm{opt}} - \mu_{\mathrm{naive}})/\sigma_{\mathrm{opt}} = 1.006$
    for symmetric mass ratio,
    where $\mu_{\mathrm{opt}}$ and $\sigma_{\mathrm{opt}}$ 
    are parameters of the optimized
    NAL models and $\mu_{\mathrm{naive}}$ is the sample mean.
Some key events have been highlighted in figure \ref{fig:nal-aligned3d_source}
    and table \ref{tab:aligned3d_source}.
As described in section \ref{sec:nal-optimization},
    the mean value of the marginal KL divergences is used
    to characterize each fit.

\end{subsection}
\begin{subsection}{Precessing spin models}

\begin{figure}
\centering
\includegraphics[width=0.48\columnwidth]{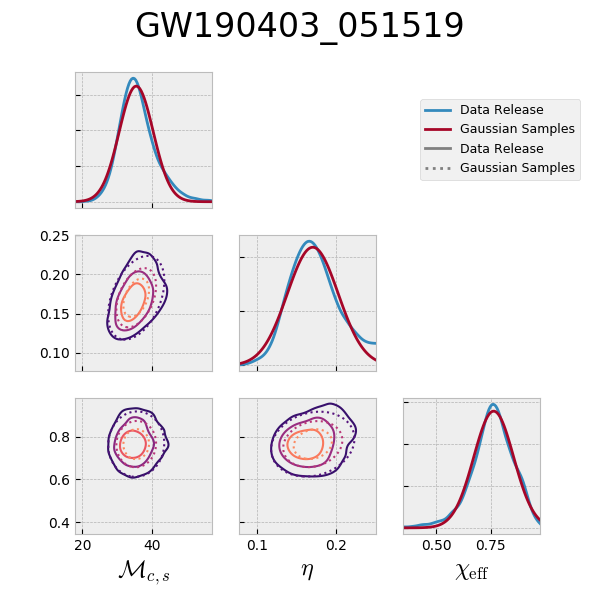}
\includegraphics[width=0.48\columnwidth]{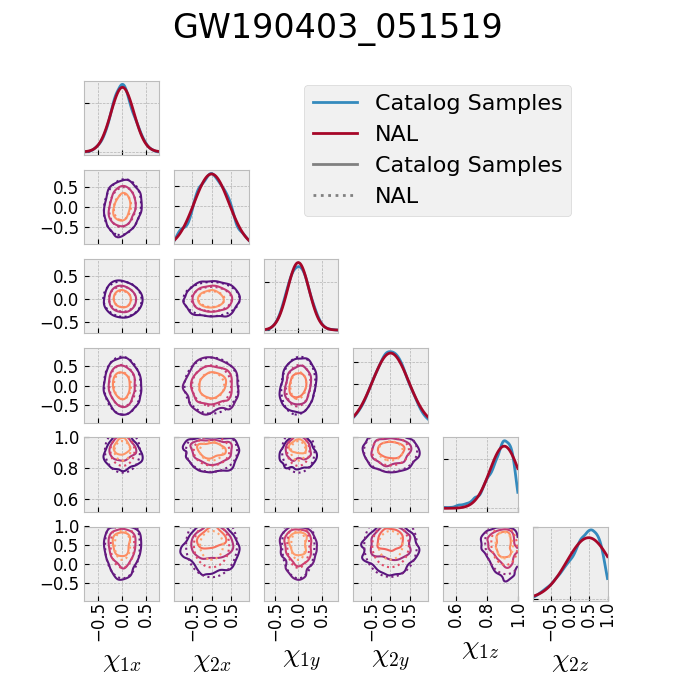}
\caption{\label{fig:nal-precessing}
These corner plots are similar to figure \ref{fig:nal-justification},
    with (left) source-frame aligned spin parameters of
    GW190403\_051519 and (right) Cartesian spin components for the same event.
\correction{We study the SEOBNRv4PHM samples for this example.}
If this marginal detection is real,
    it has a nonzero spin by 10 characteristic standard deviations.
We do note a minor disagreement in $\eta$ (left) and with
    $\chi_{1z}$ and $\chi_{2z}$ (right),
    which we attribute to non-Gaussianity in our posterior samples.
}
\end{figure}

\begin{table}
\centering
{
\small
\begin{tabular}{|l|c|c|c|c|c|}
\hline
Event & KL & $\mu_{\mathcal{M}_c}$ & $\mu_{\eta}$ &$\mu_{\chi_{1z}}$ &$\mu_{\chi_{2z}}$ \\
\hline
GW190403\_051519  & $0.009$& $36.0^{+5.4}_{-5.4}$ & $0.17^{+0.03}_{-0.03}$ & $0.91^{+0.09}_{-0.09}$ & $0.49^{+0.58}_{-0.58}$ \\
GW190517\_055101 & $0.005$ & $25.8^{+2.5}_{-2.5}$ & $0.2498^{+0.002}_{-0.003}$ & $0.66^{+0.19}_{-0.19}$ & $0.43^{+0.41}_{-0.41}$ \\
\hline
\end{tabular}
}
\caption{\label{tab:precessing8d_source}
Single valued parameter estimates for gravitaional wave events
    in the fully precessing Cartesian spin component source-frame model.
Mean KL divergences are included to demonstrate the goodness of fit.
SEOBNRv4PHM is used for GW190403\_051519,
    and ``Publication Samples'' is used for GW190517\_055101.
}
\end{table}

NAL models are in strong agreement with the GWTC samples,
    providing well-posed likelihood models for
    aligned and precessing spin.
As seen in figure \ref{fig:nal-precessing},
    NAL models overcome the finite boundary (truncation)
    effect of spin magnitudes ($\chi < 1$ for physical systems)
    for events with high spin.
While few events are known to have high spin,
    there are well-constrained examples of such events,
    and of events with precessing spin components
    \cite{GW190412, GWTC-2}.
As a system with a spin magnitude $\abs{\chi} \geq 1$ is not physical
    (representing a naked singularity in the case of a black hole),
    we must consider this another finite boundary limitation of our model.
NAL optimization overcomes this boundary effect as well, with
    $(\mu_{\mathrm{opt}} - \mu_{\mathrm{naive}})/\sigma_{\mathrm{opt}} = 0.43$
    for GW190403\_051519 and $0.28$ for GW200115\_042309 for $\chi_{1z}$,
    for example.
We provide NAL models for the 
    full six-dimensional Cartesian spin component model for each event
    where parameter estimation samples in the GWTC releases support
    such a decomposition.

\end{subsection}

\begin{subsection}{Tidal Parameters}

\begin{figure}
\centering
\includegraphics[width=0.48\columnwidth]{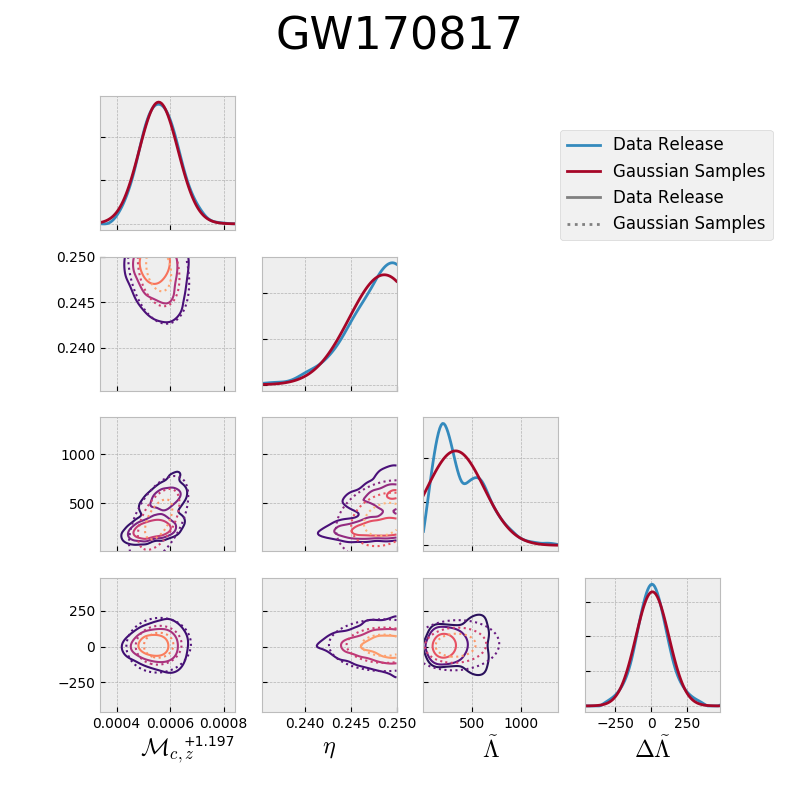}
\includegraphics[width=0.48\columnwidth]{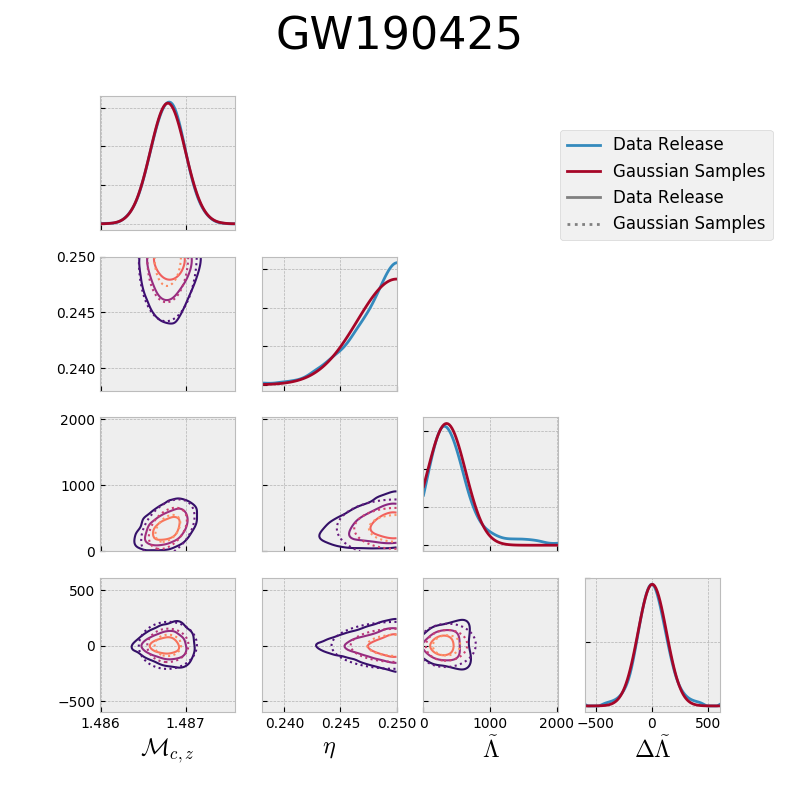}
\caption{\label{fig:nal-tides}
These corner plots are similar to figure \ref{fig:nal-justification},
    with tidal deformability parameterizations of BNS mergers
    (left) GW170817 and (right) GW190425.
\correction{
In this example in particular, we use detector-frame chirp mass
    to characterize each event.
We consider the `IMRPhenomPv2NRT\_lowSpin' samples for GW170817
    and the `AlignedSpinInspiralTidal\_LS' samples for GW190425.
We note significant non-Gaussianity in the sample distribution
    for GW170817, which explains the limitations of the NAL model
    in fully characterizing these sample distributions.
}
}
\end{figure}

\begin{table}
\centering
{
\small
\begin{tabular}{|l|c|c|c|c|c|}
\hline
Event & KL & $\mu_{\mathcal{M}_c}$ & $\mu_{\eta}$ & $\mu_{\tilde{\Lambda}}$ & $\mu_{\delta \tilde{\Lambda}}$ \\
\hline
GW170817 & $0.02$ & $1.19756^{+0.00007}_{-0.00007}$ & $0.249^{+0.001}_{-0.004}$ & $332^{+291}_{-291}$ & $9.484^{+0.011}_{-0.011}$ \\
GW190425 & $0.02$ &$1.4874^{+0.0004}_{-0.0004}$ & $0.23^{+0.02}_{-0.02}$ & $208^{+884}_{-208}$ & $37^{+192}_{-192}$ \\
\hline
\end{tabular}
}
\caption{\label{tab:aligned_tides_source}
Single valued parameter estimates for gravitaional wave events
    in the mass\_tides parameterization.
Mean KL divergences are included to demonstrate the goodness of fit.
The SEOBNRv3 approximant is used for GW150914, while the 
    IMRPhenomPv2NRT\_lowSpin approximant is used for GW170817.
}
\end{table}

As seen in figure (\ref{fig:nal-tides}),
    NAL models support tidal deformability parameterizations as well,
    and these are also in strong agreement with the GWTC samples.
We provide NAL models with tidal parameterizations
    indiscriminately for event/waveform combinations in the GWTC
    releases which offer support for tidal information,
    including the two BNS mergers, GW170817 and GW190425
    \cite{abbott2017gw170817,GW190425}.
We adopt the standard parameterization of $\tilde{\Lambda}$ and
    $\delta \tilde{\Lambda}$ from Wade et al.
    \cite{LambdaTilde}:
\begin{align}\label{eq:lambda-tilde}
    \tilde{\Lambda} = \frac{8}{13} \Big[
        (1 + 7\eta - 31\eta^2) (\Lambda_1 + \Lambda_2) +
        \sqrt{1 - 4\eta} ( 1 + 9 \eta - 11 \eta^2)(\Lambda_1 - \Lambda_2)
        \Big]
\end{align}
\begin{multline}\delta \tilde{\Lambda} = \frac{1}{2}\Big[
    \sqrt{1 - 4\eta}(1 - \frac{13272}{1319}\eta + \frac{8944}{1319}\eta^2)
        (\Lambda_1 + \Lambda_2) + \\
    (1 - \frac{15910}{1319}\eta + \frac{32850}{1319}\eta^2 + 
        \frac{3380}{1319}\eta^3) (\Lambda_1 - \Lambda_2)
\Big]
\end{multline}

\begin{figure}
\centering
\includegraphics[width=\columnwidth]{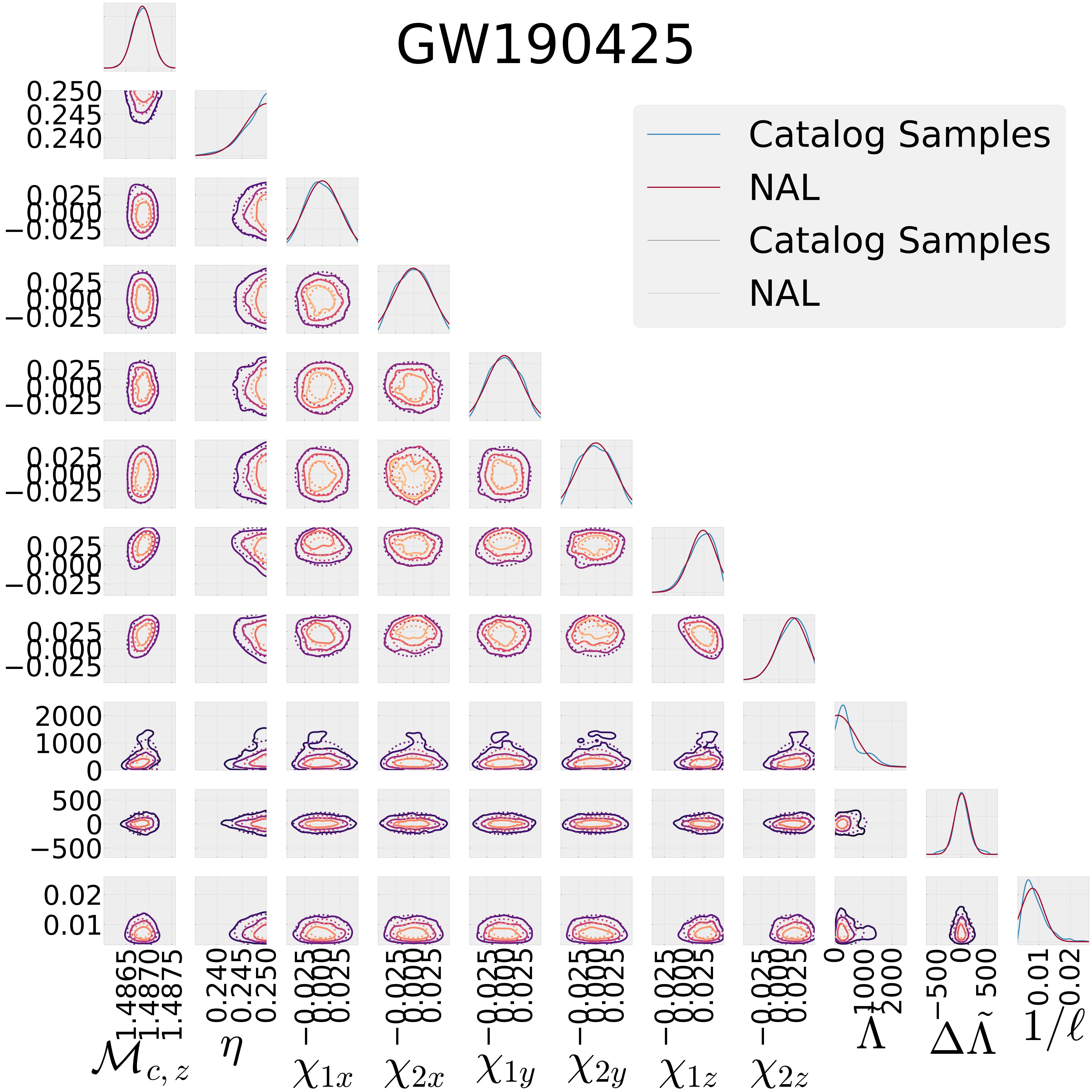}
\caption{\label{fig:nal-full-precessing-tides}
This corner plot is similar to figure \ref{fig:nal-justification},
    including the full eleven-dimensional parameterization
    for the detector-frame masses, Cartesian spin components,
    tidal deformability, and inverse luminosity distance.
\correction{
For this example, `IMRPhenomPv2\_NRTidal-LS' samples are used.
Despite some minor non-Gaussianities, the NAL model fits
    the samples well.
}
}
\end{figure}

For clarity, figure \ref{fig:nal-tides} showcases only the mass parameters of
    these fits, in detector-frame coordinates.
NAL models overcome the finite boundary effect introduced
    by truncation for $\tilde{\Lambda} > 0$.
For GW190425, for example, the boundary effect removed can be estimated by
    $(\mu_{\mathrm{opt}} - \mu_{\mathrm{naive}})/\sigma_{\mathrm{opt}} = 0.72$.
We include an example, demonstrating that NAL models apply to high-dimensional
    parameterizations of the astrophysical properties of gravitational-wave events
    in figure \ref{fig:nal-full-precessing-tides},
    exploring detector-frame mass parameters, Cartesian spin,
    tidal deformability, and inverse luminosity distance.
\end{subsection}

\end{section}

\begin{section}{Applications of NAL Models}

NAL models provide a compact likelihood approximation for 
    the astrophysical properties of each gravitational-wave event
    published in the GWTC releases.
They further provide insight into those astrophysical properties for individual
    events through MLE characterizations of the parameters representing 
    those properties.
They also overcome bias introduced when
    assuming a Gaussian from the sample mean and covariance.
Together, these models further provide an understanding of the
    population of gravitational-wave events released so far,
    and our methods have broader applications such as population inference
    (as we will see in the next few chapters) and parameter estimation.

\begin{subsection}{Properties of the Gravitational-Wave Population}

\begin{figure}
\centering
\includegraphics[width=0.6\columnwidth]{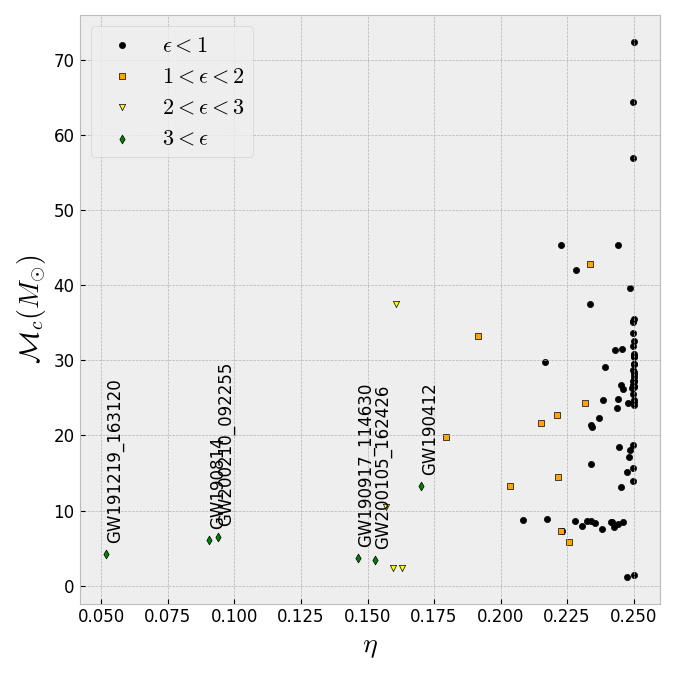}
\caption{\label{fig:nal-scatter}
MLE aligned-spin source-frame estimates in $\mathcal{M}_c$ and $\eta$ for all
    marginal and confident detections in the GWTC releases.
Markers indicate deviation from equal mass
    ($\epsilon=(1/4 - \mu_{\mathrm{opt}})/\sigma_{\mathrm{opt}}$).
}
\end{figure}

Through GWTC-3,
    the gravitational-wave population is reported
    to be biased toward equal mass \cite{LIGO-O3-O3b-RP}.
However, the sample-mean estimates tabulated for the properties of each event
    in the GWTC papers don't provide an effective
    MLE representation of the symmetric mass ratio,
    which waveform models are especially sensitive to.
Figure \ref{fig:nal-scatter}
    demonstrates the NAL maximum likelihood estimates of
    each event (marginal and confident),
    in source-frame chirp mass and symmetric mass ratio.
Our parameterizations of each likelihood model
    allow us to estimate how many $\sigma$ each event
    is in difference from equal mass.
This quantity is related to the half-width-half-max of our distribution,
    and is a characteristic estimate of the standard deviation
    expected of the Gaussian without a boundary
    (used to classify the error of measurements as 
    $\sigma$, $2\sigma$, $3\sigma$ etc...).

\begin{figure}
\centering
\includegraphics[width=0.6\columnwidth]{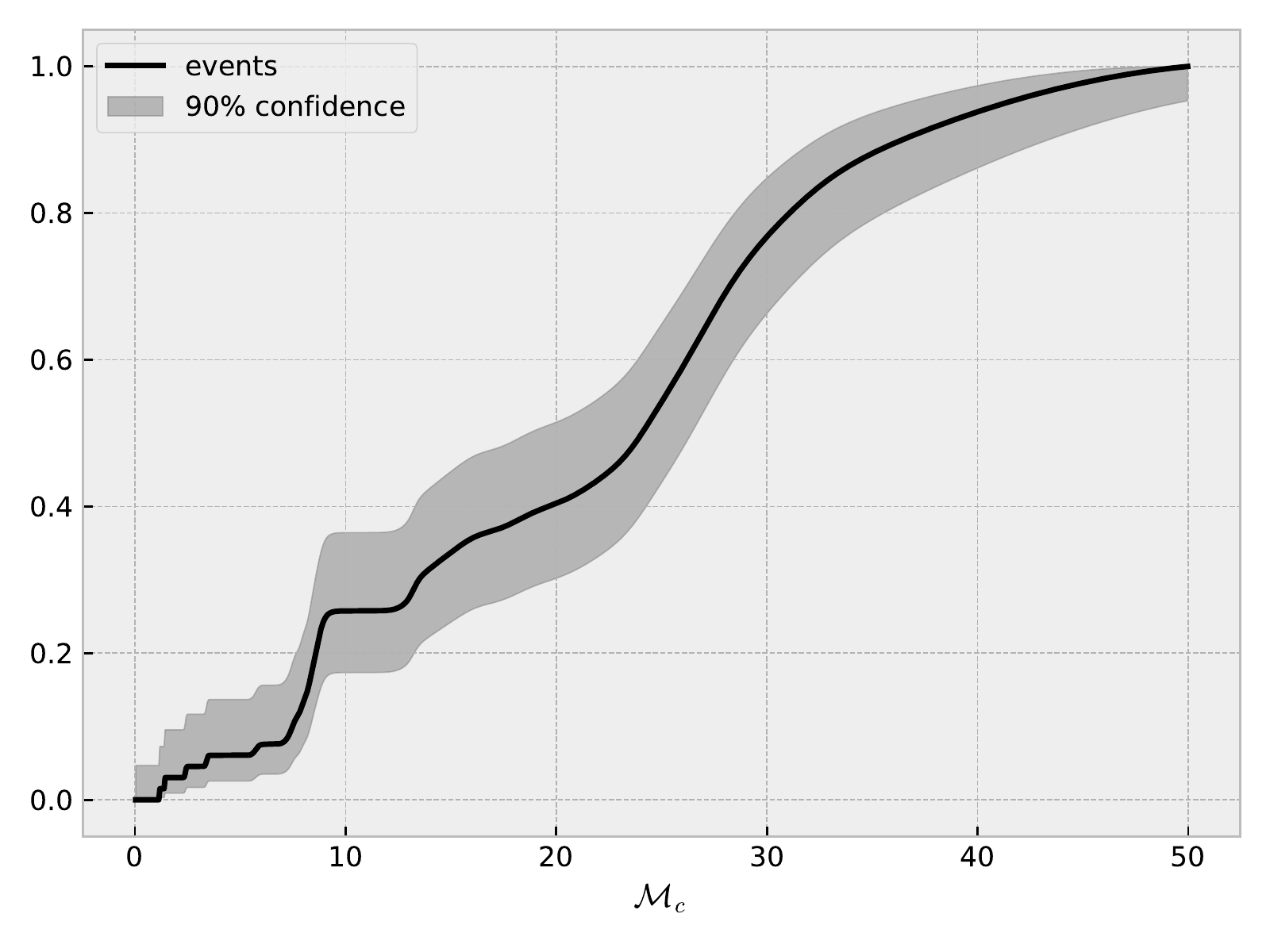}
\caption{\label{fig:nal-cdf}
The CDF for the gravitational-wave population,
    estimated using NAL models with a 
    \correction{Wilson Score interval} \cite{WilsonScore},
    including only those events in the LSC's O3b population paper
    \cite{LIGO-O3-O3b-RP}.
}
\end{figure}

In addition to MLE characterizations,
    we can also describe the population using the truncated Gaussians
    in various ways.
For example, figure \ref{fig:nal-cdf} arranges the
    Cumulative Distribution Function (CDF)
    for the gravitational-wave population through GWTC-3,
    using only the confident detections
    (consistent with \cite{LIGO-O3-O3b-RP}).
For this arrangement, the one-dimensional marginal
    for the source-frame chirp mass is used
    for a variety of approximants.

\end{subsection}
\begin{subsection}{Applications for Low-Latency Parameter Estimation}

\begin{figure}
\centering
\includegraphics[width=0.6\columnwidth]{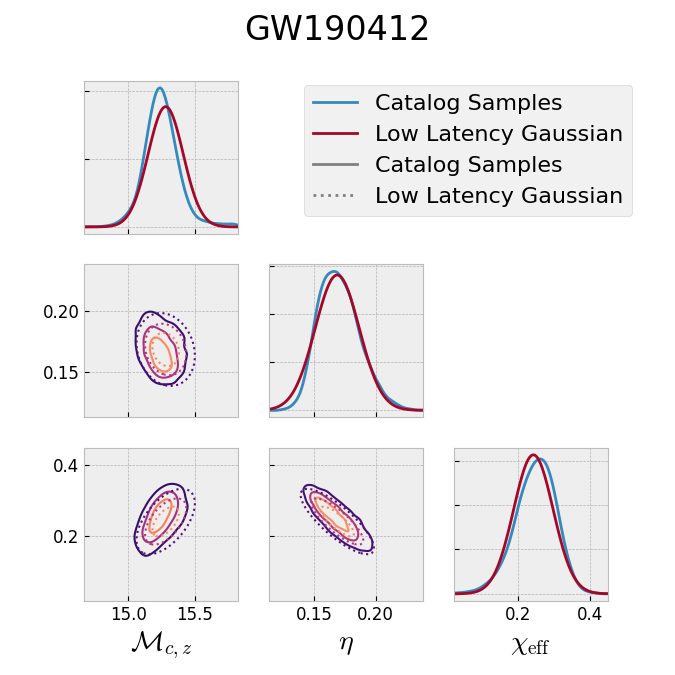}
\caption{\label{fig:nal-lowlatency}
A NAL (detector-frame) characterization of GW190412,
    using a sparse set of likelihood evaluations from RIFT
    (with IMRPhenomD),
    designed to imitate the availability of evaluations
    in a low-latency setting.
The GWTC sample estimate provided for comparison
    are the ``PublicationSamples'' provided in the GWTC-2 release.
}
\end{figure}

Our method can ultimately be applied not only to
    the secondary parameterization
    of released samples, but also directly to parameter estimation
    routines.
The NAL optimization can be carried out using either
    a set of draws from the posterior of a given event
    (as done throughout this chapter)
    or a set of likelihood evaluations.
Even a sparse set of evaluations can be used to efficiently generate
    NAL models,
    making NAL a candidate for the rapid characterization
    of binary sources in-situ during an observing run.

Figure \ref{fig:nal-lowlatency} provides a simplistic example of this potential
    low-latency application.
The RIFT inputs provide a limited set of likelihood evaluations
    through IMRPhenomD, using marginalized detector-frame
    mass and spin distributions.
This evaluation set is chosen in the range of
    binary parameters flagged during an observing run,
    for low-latency followup for a potential electromagnetic counterpart.
As most potential electromagnetic follow-ups in the near future
    are expected to occur at low redshift,
    a detector-frame mass parameterization is employed
    in this example.
\correction{
The optimization routine for this example is simplified
    so as to not select parameters for the \textit{child} population
    using information about the one- and two-dimensional marginals,
    as these are not computed for a grid of likelihood evaluations.
}
The NAL optimization in this example is carried out in a few seconds,
    including the overhead costs of initializing the program.

The exceptional, uninformed, agreement of the low-latency NAL 
    optimization (figure \ref{fig:nal-lowlatency})
    with the published GWTC samples
    demonstrates the possibilities our method holds
    for low-latency parameter estimation application.
However, further work must be done for a systematic justification
    of this method for potential low-latency parameter estimation.

\end{subsection}
\end{section}

 \end{chapter}

\begin{chapter}{Simulating the Evolution of Massive Binary Star Systems}
\label{chap:binary}
In this chapter, I provide a brief overview of the binary evolution
    of massive stars, focusing on the processes which 
    are characterized in this dissertation and related work.
I will begin with a brief review of the literature for massive
    stellar binaries.
Following this, I will describe our use of the StarTrack algorithm
    used to produce the population of simulated mergers associated with
    the formation channel for compact binaries.
I will demonstrate the processing required to construct 
    a physically justified population of merging binaries
    to be consistent with those observed by gravitational-wave
    observatories.
Finally, this chapter will also include a discussion of the StarTrack
    synthetic universe data products, as well as their physical significance.

\begin{section}{The Evolution of Massive Stars and Binaries}
\label{sec:lit-review}

We summarize the well-studied life cycle of massive stars;
    see Hurley et al. (2000) for a review
    \cite{Hurley2000}.
See also other related works
    \cite{book-stars-Clayton,book-stars-KW,
    WoosleyReview2002}.
When a star is born, it will reach a thermodynamic equilibrium
    with the nuclear reactions that power it;
    i.e. the fusion of hydrogen into helium.
Once this equilibrium is achieved, we say that a star has
    reached the Zero Age Main Sequence (ZAMS),
    where its mass, metallicity, and rotation
    become sufficient to determine a star's isolated evolution.
This is the starting point for the study of stellar evolution.

\correction{
A star spends most of its life on the main sequence,
    fusing hydrogen in its core.
Following this, the course of a star's evolution
    depends largely on its mass.
The least massive stars ($~0.08 M_{\odot} < m < ~0.8 M_{\odot}$)
    never fully expend the hydrogen at their cores,
    and continue to burn hydrogen slowly,
    conserving their fuel for potentially trillions of years.
Stars in the ($~0.8 M_{\odot} < m < ~ 8 M_{\odot}$) 
    will have a shorter lifetime,
    entering the red giant and asymptotic giant branch stages
    before ending their lives as white dwarfs
    with planetary nebula.
}

\correction{
A massive star ($M > ~8 M_{\odot}$) will leave the main sequence
    even sooner, as a supergiant.
As giant and supergiant stars cool and expand,
    they lose a substantial amount of mass to stellar wind.
They } may also shed the outer layers of its atmosphere \correction{entirely},
    becoming an exposed helium core or Wolf-Rayet star.
Further nuclear reactions are endothermic, and the star's
    core will collapse as electron degeneracy pressure is lost.
The end of such star's life is a type II supernova explosion,
    the result of which is a compact remnant
    (a neutron star or a black hole).

It is expected that about two thirds of massive stars
    belong to a binary system \cite{Belczynski2002}.
These binary systems are bound together by the gravitational
    binding energy of their orbit and separated
    by the orbital angular momentum.
For those systems of massive stellar binaries in tightly bound orbits,
    several of effects can determine if they remain separated
    for their lifetime, or if enter a phase of mass transfer (accretion),
    enter a phase where they share a common envelope of external atmosphere,
    or merge as luminous companions or compact remnants with 
    a detectable gravitational radiation signature.
The rest of section \ref{sec:lit-review} will explore how
    these processes are modeled,
    in order to predict how stars and stellar binaries evolve.

\begin{subsubsection}{Stellar Wind}

Stellar wind is material ejected from a stellar atmosphere
    as gas and dust without substantial clumping.
\correction{
Stellar wind is key to the volumetric mass loss rate of stars.
Understanding it requires a study of stellar atmosphere models.
}
Various groups have studied mass loss due to stellar wind in the evolution
    of massive stars
    \cite{BoffinJorissen1988, Nieuwenhuijzen1990, Hamann1995, Mink2001,
        Oskinova2011, Brott2011, Meynet2015, Ramachandran2019}.
In addition to its role in stellar evolution,
    some stellar wind can be captured by a binary companion,
    effecting binary evolution
    \cite{BoffinJorissen1988,Belczynski2008,Belczynski2020b}.

The initial mass and metallicity of a star are expected to play a major role in
    the amount of matter ejected by stellar wind,
    with substantially weaker stellar winds present in main sequence stars
    at low metallicity
    \cite{Oskinova2011,Ramachandran2019,Belczynski2020b}.
This is known as the ``weak-wind phenomenon,'' which is an ongoing area
    of research, and may determine if a star enters a Wolf-Rayet phase
    or not \cite{Ramachandran2019}.
However, this phenomenon does not substantially impact
    volumetric mass loss rates during
    the supergiant phase of massive stars \cite{Ramachandran2019}.

\end{subsubsection}

\begin{subsection}{Supernova Engines}
\label{sec:supernova}

The death of a massive star through a core-collapse supernova
    has been studied in-depth by many groups
    \cite{Fryer2012, Nordhaus2006, Leung2019},
    and we attempt to briefly characterize the supernova mechanisms.
It is important to understand these processes
    when studying the formation channels for compact binaries,
    as different models are capable of producing
    black holes and neutron stars in different mass ranges,
    and with different properties.
It was common knowledge that black holes should not form
    in the lower mass gap (2.5 -5 $M_{\odot}$),
    in a vacuum, until gravitational-wave events were observed
    with masses in that range (e.g. GW190814;
    although it was proposed that black holes could form in that range
    due to interactions with a binary companion of similar 
    ZAMS mass) \cite{GW190814}.

As the runaway nuclear reactions in massive stars become endothermic,
    the core of a massive star begins to collapse.
This collapse is halted by nuclear forces,
    as a proto-neutron star core forms with the density of an atomic nucleus.
Infalling material during this collapse can have velocities
    near the speed of light,
    and the formation of the proto-neutron star core halts this
    infalling material,
    producing shocks that reverberate throughout the convective
    envelope of the star.
The proto-neutron star continues to leak energy through
    neutrino emission, as neutrinos are particles which carry mass
    and can escape from dense regions where light is blocked by
    the stellar atmosphere.
The shocks carry energy away from the core until they are damped by neutrino
    emission.
Instabilities in the region where these shocks propagate can
    result in energy instead being transformed into kinetic energy,
    pushing the envelope away from the core.
This kinetic energy is the cause of the energetic ejection of material
    associated with the supernova.
The time taken by the ejection of the convective envelope determines
    the amount of matter which can accrete onto the proto-neutron star core,
    and ultimately determines the fate of the compact remnant.

Groups such as Fryer et al. have proposed rapid and delayed
    convection-enhanced neutrino-driven supernova explosion models
    \cite{Fryer2012}.
The ``rapid'' supernova engine predicts a lower mass gap for black holes,
    as stars undergoing a core-collapse supernova with
    a ``rapid'' convective timescale are unable to produce compact remnants
    in the $2.5-5 M_{\odot}$ range.
The ``delayed'' supernova engine does not exhibit the same properties
    and is capable of producing black holes in this region
    \cite{Fryer2012, Belczynski2020}.

There are other types of supernovae which can occur for the most massive
    stars ($M > ~80 M_{\odot}$) \cite{Leung2019}.
For these stars,
    energetic gamma rays are produced during the formation of the
    oxygen core, the absorption of which causes
    electron-positron pair production.
This causes a contraction of the core, resulting in an accelerated
    oxygen burning phase.
For stars with more than $150 M_{\odot}$,
    this accelerated oxygen burning becomes a runaway thermonuclear explosion,
    annihlilating the star in a Pair-Instability SuperNovae (PISN),
    leaving no compact remnant.
For stars in the $80 M_{\odot} < M < 150 M_{\odot}$ range,
    the energy released by pair-instability is insufficient
    to trigger a PISN.
These stars release energetic pulses during the pair-instability phase,
    shedding mass until they no longer produce the energetic 
    gamma rays required for pair-instability,
    and ultimately undergo core-collapse supernovae.
Systems which undergo this Pulsational Pair Instability (PPI)
    before core-collapse are labeled as PPI SuperNovae (PPISNe).

The effect of PISN and PPISN on the population of compact remnants
    is the prediction of an upper limit on low-mass black holes,
    around $50 M_{\odot}$, known as the ``upper mass gap.''
The mass ranges for which PISN and PPISN can occur
    depend on the rotation and
    the ZAMS metallicity of massive stars,
    as well as models of mass loss due to stellar wind
    \cite{Leung2019}.

\begin{subsubsection}{Supernova Recoil Kicks}
Asymmetries in ejected material during
    a supernova can impart linear momentum on 
    a compact remnant, which results in a ``recoil kick''
    for the compact remnant, with some velocity.
The strongest evidence for supernova recoil kicks
    is that the velocity distribution of neutron
    stars is much greater than that of the stellar populations
    which produce them
    \cite{Gunn1970, Chatterjee_2005, Nordhaus2006}.
Observationally, there is further support for recoil kicks
    in both gravitational-wave measurements \cite{OShaughnessy2017}
    and electromagnetic sources \cite{Gualandris_2005, Poutanen2022}.
For neutron stars, 
    this effect is equal to the linear momentum of the ejected material
    \cite{CordesI, Hobbs2005}.
Belczynski et al. (2002) found that a second Maxwellian component
    is a good predictor for pulsars with high kick velocities
    \cite{Belczynski2002}.
For black holes, the imparted linear momentum is reduced by the fraction
    of the ejected material that falls back onto the black hole
    \cite{CordesII, Fryer2012}.
Recently, different groups have found support for substantial asymmetries
    in supernova engines driven by neutrinos,
    using independent simulation and analysis
    \cite{Wongwathanarat2010, Fryer2012, Nordhaus2012}.

\end{subsubsection}

\end{subsection}
\begin{subsection}{Interacting Binaries and the Common Envelope Phase}

Different mechanisms are expected to enable mass transfer
    between binary companions as well as mass ejection
    from binary systems,
    including the Common Envelope (CE) phase as well as 
    stellar wind, clumped ejection, and other means.
These interactions typically involve the transfer of mass from
    one star (the donor) to its companion (the accretor).
The most significant of these interactions is the ejection
    of a common envelope by means of non-conservative mass transfer,
    which is essential to our understanding of how compact objects form
    \cite{BondiHoyle1944, Paczynski1976,
        Rappaport1982, Rappaport1983,
        Livio1988, Podsiadlowski1992, Belczynski2002}.

Common envelope evolution can result in 
    compact binaries, 
    cataclysmic variable stars,
    Type Ib supernova, naked helium stars,
    or a complete merger without the formation of a compact binary
    \cite{Paczynski1976,Klencki2021}.
In particular, the formation channel for compact binaries which
    may merge to produce gravitational radiation involves
    a rapid inspiral of the companion object
    inside the shared envelope, which leads to the ejection of the common
    envelope.
There are not many other known mechanisms by which
    compact binaries may experience the necessary orbital 
    shrinkage to merge by emission of gravitational radiation
    within the age of the known Universe.
Conversely, if the common envelope remains bound,
    the binary components merge completely without forming a compact binary.

A successful CE ejection occurs when the remnant core of the giant
    donor star is smaller than its Roche-lobe, 
    and depends largely on the original mass, radius, and metallicity
    of the donor star, as well as a choice of common envelope efficiency
    $\alpha_{CE}$, and an envelope structure constant $\lambda_{CE}$
    \cite{XuLi2010, 
        Briel2022, Fryer1999, Moreno2021, Zwart1998, Tauris1996, Zwart1995,
        Hachisu1999,
        IvanovaEnthalpy2011, IvanovaRemnant2011, Wang2016,
        Klencki2021, Belczynski2008, Belczynski2022,
        broekgaarden2021formation, Wilson2022}.
The energy stored in a common envelope can be calculated by considering
    the gravitational potential energy of the envelope 
    together with thermal and chemical energies.
By modeling the gravitational potential energy of the common envelope
    and core of the donor star separately,
    we can calculate the post-CE separation of a binary
    while freely varying both $\lambda_{CE}$ and $\alpha_{CE}$
    \cite{Klencki2021}.

Mass transfer instability plays a crucial role in this process,
    affecting the structure of the common envelope
    as well as the distance between the core of each companion.
Mass transfer becomes unstable when the donor star's Roche-lobe
    shrinks faster than the size of the donor,
    and is the result of a non-conservative
    method of mass transfer on behalf of the donor
    \cite{Podsiadlowski1992}.
This type of non-conservative mass transfer
    necessarily occurs only in systems with a mass
    ratio satisfying $q_{crit} < q < 1.0$, for some $q_{crit}$
    \cite{Podsiadlowski1992}.
Rather than a pre-determined constant of our evolution model,
    $q_{crit}$ is constrained by the population of merging binaries.
If mass transfer occurs after the 
    hydrogen core burning phase, but before the ignition of
    central helium, and the primary fills its Roche lobe,
    then $q_{crit} \approx 0.97$
\cite{AbtLevy1978, TutukovYungelson1980CloseBinaryStars}.
Alternatively, if mass transfer occurs after the initial 
    primary's helium burning phase, $q_{crit} \approx 0.9$.
Supernova recoil kick velocities have a large impact on the outcome of such
    mass transfer associated with a common envelope ejection,
    as radially symmetric supernovae are more likely to cause
    a binary system to become unbound
    \cite{Bailes1989}.

Here, we describe the formalism for this type of non-conservative mass
    transfer, as seen in Rappaport,
    Joss, and Webbink (1982), and Rappaport, Verbunt, and Joss (1983)
    \cite{Rappaport1982, Rappaport1983}.
This process is specified by two free parameters:
    The fraction $\STFa$, of the mass lost by the doner, that is accreted
    by star 2, and the specific angular momentum $\STbeta$, of any matter
    lost from the system, in units of $2\pi a^2/P$.
\begin{equation}
\delta J = \STbeta \delta \dot{M_1} (1 - \STFa) \frac{2\pi a^2}{P}
\end{equation}
where $\delta \dot{M_1}$ is infinitesimal mass lost by the initial primary,
    and $\delta J$ is a small amount of angular momentum held by matter
    ejected by the system, $a$ is the semimajor axis, and $P$ is the orbital
    period
    \cite{Podsiadlowski1992}.
Values of $\STFa$ less than unity indicate mass is lost from the system
    (due to ejected material).
Podsiadlowski et al (1992) \cite{Podsiadlowski1992}
    argue that if the common envelope is in thermal equilibrium,
    the final properties of the binaries depend only on the
    final masses of the hydrogen-exhausted core and hydrogen-rich
    envelope.

\end{subsection}
\end{section}

\begin{section}{Simulating Populations of Massive Binaries with StarTrack}

Many groups have developed binary evolution simulations to produce 
    synthetic populations of compact binaries,
    including the COMPAS software \cite{StevensonCOMPASNature2017,
    RileyCOMPAS2022, broekgaarden2021formation,
    broekgaarden2021impact}
    and POSYDON \cite{posydon}.
The POSYDON group (Fragos et al) \cite{posydon} have organized a list
    including many of these other software
    \cite{
    Izzard2004, Izzard2006, Izzard2009,
    Eldridge2017,
    Vanbeveren1998a, Vanbeveren1998b,
    Hurley2002,
    Kruckow2018,
    Breivik2020,
    Giacobbo2017,
    Lipunov2009,
    Spera2015,
    Toonen2012, Toonen2016}.
Throughout the rest of this dissertation,
    the work summarized includes the StarTrack binary evolution simulations
    \cite{Belczynski2002, Belczynski2008, Belczynski2016Nature, Belczynski2020}.

The StarTrack algorithm reproduces
    a physically motivated population of massive stellar binaries
    for a population with a specified metallicity and initial
    mass function.
The merger rate densities represented by these populations
    can be used to generate synthetic universes which predict 
    the compact binaries formed in the history of The Universe,
    accounting for changes in the metallicity of The Universe over
    cosmological time \cite{DominikII}.
StarTrack accounts for many physical processes, including
    stellar evolution, accretion, tidal interactions,
    stellar wind, metallicity, gravitational radiation,
    magnetic braking, and recoil kicks.
The content in this section is based on prior work
    \cite{Belczynski2002, Belczynski2008, Belczynski2016Nature, Belczynski2020,
        DominikI, DominikII, DominikIII, Wysocki2019}.

Each run of StarTrack begins with
a set of assumptions about the physical processes involved
    in the formation and evolution of binary star systems which
    could produce compact binaries.
In many cases, these assumptions are characterized by
    a formation parameter assumed inductively.
The set of these formation parameters defines the independent variable
    space of our simulations, which we call $\Lambda$.

\begin{subsection}{Initial Mass Function}

A StarTrack synthetic universe is generated as a set of distinct
    runs of the core StarTrack algorithm,
    each corresponding to a cosmological reference time.
Each such run is generated as a fixed number of isolated binaries
    at a fixed reference time.
The primary component of each binary is drawn from 
    \correction{a Kroupa}
    Initial Mass Function (IMF)
    \cite{Salpeter, kroupa1993, kroupa2003}.
\begin{equation}\label{eq:IMF}
\Psi(M_1) \propto \begin{Bmatrix}
                  M^{-1.3}_1 & 0.08 M_{\odot} \leq M_{1} < 0.5 M_{\odot} \\
                  M^{-2.2}_1 & 0.5 M_{\odot} \leq M_{1} < 1.0 M_{\odot} \\
                  M^{-\alpha_IMF}_1 & 1.0 M_{\odot} \leq M_1 < 150 M_{\odot} \\
                  \end{Bmatrix}
\end{equation}
where we adopt $\alpha_{IMF} = 2.35$, consistent with the literature
    \cite{Belczynski2020,Klencki2018}.
The companion mass is drawn from a uniform distribution in mass ratio,
    for $q_{\mathrm{min}} \leq q \leq 1.$
    \cite{kobulnicky, Bastian2010, Duchene2013Review}.
A value of $q_{\mathrm{min}} = 0.08 M_{\odot}/m_1$ represents the 
    smallest allowed companion corresponding to the hydrogen-burning limit
    \cite{kobulnicky}.
Following Belczynski et al. (2020) \cite{Belczynski2020},
    we adopt a value for $q_{\mathrm{min}} = 0.1 M_{\odot}/m_1$.
Only systems for which $m_1 > 5 M_{\odot}$ and $m_2 > 3 M_{\odot}$
    are evolved in a StarTrack run, as only these systems
    are expected to form compact binaries.

This population of stars is representative of some amount of star-forming gas,
    $M_{\mathrm{sim}}$,
    which must account for systems in the mass cutoff region,
    systems which do not form binaries,
    and systems which do not merge in the age of The Universe.
The contribution for systems outside the mass cutoff for potential 
    compact binaries must be described by a mass efficiency,
    $\lambda_{\mathrm{eff}}$,
    \cite{Richard2010}.
\begin{equation}\label{eq:mass-frac}
\lambda_{\mathrm{eff}} = \frac{n}{N} \frac{f_{cut}}{\langle M \rangle}
\end{equation}
where $N$ binaries are simulated out of $n$ progenitor systems,
    $\langle M \rangle$ is the average mass of all binary progenitors,
    and where $f_{\mathrm{cut}}$ accounts for the proportion of the IMF
    in the cutoff region.

The true fraction of stars, $f_{\mathrm{bin}}$,
    which belong to binary systems is not known.
We compare the merger rate per unit mass at a given time,
    marginalized over merger parameters,
    $\rho_{\Lambda}(t|f_{\mathrm{bin}})$,
    to a reference value where $f_{\mathrm{bin}} = 1$.
This way, we can rescale the merger rate for an arbitrary choice
    of $f_{\mathrm{bin}}$ \cite{Richard2010}:
\begin{equation}\label{eq:fbin}
\rho_{\Lambda}(t|f_{\mathrm{bin}}) = 
    \rho_{\Lambda}(t|f_{\mathrm{bin}} = 1)
    \frac{f_{\mathrm{bin}}(1 + \langle q \rangle)}{
        1 + f_{\mathrm{bin}} \langle q \rangle}
\end{equation}
For our purposes, $\langle q \rangle$ is assumed to be one, and
    $f_{\mathrm{bin}}$ is assumed $1/2$. 

Finally, we can compute $M_{\mathrm{sim}}$, the mass represented
    by the represented population of StarTrack binaries:
\begin{equation}\label{eq:msim}
M_{\mathrm{sim}} = \frac{2f_{\mathrm{bin}}}{1 + f_{\mathrm{bin}}}
    \frac{n}{N}
    \frac{f_{\mathrm{cut}}}{\langle M \rangle}
    M_{\mathrm{bin}}
\end{equation}
This mass estimate allows us to rescale the population of 
    binary systems evolved by the StarTrack algorithm 
    to match the size of a galaxy, or the physical universe.
Each metallicity bin evolves $N = 2 \times 10^{6}$ binary systems
    \cite{Belczynski2020}.

\end{subsection}

\begin{subsection}{Single Star Evolution Assumptions}

StarTrack relies extensively on the work of Hurley, Pols, \& Tout (2000),
    for the evolution of massive stars,
    summarized in section \ref{sec:lit-review} \cite{Hurley2000}.
Individual stars evolve following
    from Hurley, Pols, \& Tout (2000) \cite{Hurley2000},
    whereby an individual star's zero-age main sequence 
    mass and metallicity are sufficient to predict its evolution.
Consistent with Belczynski et al. (2020) \cite{Belczynski2020},
    we explore a variety of supernova engine settings,
    based on the work of Fryer et al. (2012) \cite{Fryer2012}
    and Leung, Nomoto, and Blinnikov (2019)  \cite{Leung2019},
    including rapid and delayed models which account for
    pair-instability.
We explore a parameter space in $\STkick$,
    the characteristic recoil kick velocity for 
    a single Maxwellian distribution representative
    of asymmetric neutrino-driven supernova ejecta.

\begin{subsubsection}{Stellar Wind}

The StarTrack algorithm models stellar wind as an accretion process,
    following Bondi \& Hoyle (1944) and Boffin \& Jorissen (1988)
    \cite{BondiHoyle1944, BoffinJorissen1988, Belczynski2008}.
To address the weak-wind phenomenon, StarTrack makes use of global
    scale factors of $\STwind \leq 1.0$, and $\STwindHe \leq 1.0$,
    which multiplies the assumed volumetric mass loss rate due to stellar wind
    \cite{Belczynski2020b}.
Here, $\STwind$ applies to hydrogen-dominated stars,
    while $\STwindHe$ applies to helium-dominated stars
    (such as naked helium cores).
In our work, we vary these factors together ($\STwind = \STwindHe$),
    but this is not a necessary assumption.

\end{subsubsection}
\end{subsection}

\begin{subsection}{Binary Evolution}

The StarTrack binary evolution algorithm considers not only
    individual stellar evolution, but also an up-to-date
    model of the effects of stellar evolution on binaries,
    consistent with literature
    (see section \ref{sec:lit-review})
    \cite{Belczynski2020}.
StarTrack makes standard assumptions about the initial 
    separation of binary systems \cite{Abt1983}.
As we are interested in only systems which can become 
    merging compact binaries, the evolution of systems
    where the donor mass falls below $0.5 M_{\odot}$, or
    when the companion falls below $0.3 M_{\odot}$ during mass
    transfer is discontinued.

\begin{subsubsection}{Formation Parameters}\label{sec:st-formation-params}

In our own work, we use StarTrack simulations like those
    described in Belczynski et al (2020) \cite{Belczynski2020}.
The models we study are variations with particular choices 
    for some of the formation parameters assumed 
    as inputs for the StarTrack simulations.
For instance, we vary our assumptions in 
    the accretion fraction $\STFa$ and specific angular momentum $\STbeta$
    associated with non-conservative mass transfer
    during common envelope evolution 
    (see \ref{sec:lit-review}).

Consistent with Belczynski et al (2020) \cite{Belczynski2020},
    the binding energy parameters $\alpha_{CE}$ and $\lambda_{CE}$
    are set constants where $\alpha_{CE} = 1$ and $\lambda_{CE} = 0.1$.
\correction{
The value $\alpha_{CE}$ is consistent with prior literature,
    and is supported by modern work
    \cite{Briel2022, Fryer1999, Moreno2021, Zwart1998, Tauris1996, Zwart1995,
        Hachisu1999, Belczynski2020,Wilson2022}.
The value of the structure constant is expected to be consistent 
    with more massive stars,
    such as those which can form compact binaries
    \cite{XuLi2010,Wang2016,Belczynski2020}.
}

\end{subsubsection}
\begin{subsubsection}{Spinning Binaries}

Previous work has indicated that the final spin of
    compact binary mergers originating from isolated binary
    evolution does not depend on initial stellar
    rotation or metallicity
    \cite{Fuller2019, MaFuller2019, FullerMa2019, Belczynski2020}.
The final spin of such a compact binary is thought to depend more on the
    process by which angular momentum is transported within the
    system.
Belczynski et al. have done recent studies on spin distributions in
    the population of compact binaries generated by the StarTrack
    binary evolution simulations \cite{Belczynski2020}.
In this work, we do not consider these spin models.

\end{subsubsection}
\end{subsection}

\end{section}

\begin{section}{Cosmological Postprocessing}

As stated previously, a synthetic universe is not one run of StarTrack,
    but several, at different metallicities
    \cite{DominikII,Belczynski2020}.
Each run of StarTrack is at a fixed metallicity, and
    simulates an epoch of steady star formation, $\Delta t$
    in cosmological time.
We now summarize the work of Belczynski et al. \cite{Belczynski2020},
    to describe the synthesis of a synthetic universe using 
    multiple StarTrack runs at different assumed metallicities.

\begin{subsection}{Estimating a Merger Rate}
\label{sec:s_i}

Each run of StarTrack outputs a set of sample mergers,
    occupying a compact binary merger parameter space,
    $\BinaryParameters$, which will have some associated density
    $\rho_{\Lambda, \metallicity}(\BinaryParameters)$,
    where $\Lambda$ is the set of physical assumptions about the
    formation parameters of merging compact binaries,
    and $\metallicity$ is the assumed metallicity.
When desired, this density is normalized over the population
    to describe a dimensionless probability distribution,
    characterized by $N_{\Lambda, \metallicity}$: 
    the integrated value of the physically motivated density
    over the entire sample.
\begin{equation}\label{eq:Nzlambda}
N_{\Lambda,\metallicity} = \int\limits_{\BinaryParameters} 
    \rho_{\Lambda, \metallicity} (\BinaryParameters) \mathrm{d} \BinaryParameters
\end{equation}

We now follow along the outlined procedures described
    in section 2.6 of Belczynski et al. (2020) \cite{Belczynski2020}.
Therefore, we adopt the Madau \& Fragos (2017)
    star formation rate variability on redshift ($\redshift$),
    with an IMF-dependent correction factor
    \cite{Madau2017}:
\begin{equation}\label{eq:sfrz}
\mathrm{SFR}(\redshift) = 
    \mathcal{K}_{\mathrm{IMF}} 0.0015
    \frac{(1.0 + \redshift)^{2.7}}{1.0 + ((1.0 + \redshift)/3.0)^{5.35}}
    \mathrm{M}_{\odot} \mathrm{Mpc}^{-3} \mathrm{yr}^{-1}
\end{equation}
Here, $K_{\mathrm{IMF}} \correction{= 0.66}$ is a correction factor, which
    readjusts the star formation rate to match the Salpeter IMF.
We also adopt the Madau \& Fragos mass-metallicity
    relationship, satisfying
    $\log (\metallicity/\metallicity_{\odot}) = 0.153 - 0.074 \redshift^{1.34}$,
    where
    \correction{$\metallicity_{\odot} = 0.017$.}

Each chunk $\Delta t$ is characterized by a sample of galaxies from
    a Schechter-type probability density function, 
    \correction{frozen at redshift $\redshift=4$,}
    including $10^4$ such galaxies
    \cite{schechter, fontana2006, DominikII}.
Each galaxy is assigned a metallicity, according to
    a relationship assumed for the dependency of metallicity on redshift
    \correction{(see \cite{Madau2017}),
    with 0.5 dex simulated Gaussian noise.}
These galaxies are gathered into metallicity bins,
    and the mass of each bin is calculated, 
    $M_{\Lambda,\metallicity,\Delta t}$.
The metallicity bins used in our studies are $\metallicity \in \{$
    $0.0001$, $0.0002$, $0.0003$, $0.0004$, $0.0005$, $0.0006$, $0.0007$, $0.0008$, $0.0009$,
    $0.001$, $0.0015$, $0.002$, $0.0025$, $0.003$, $0.0035$, $0.004$, $0.0045$, $0.005$,
    \correction{$0.0055$, $0.006$,}
    $0.0065$, $0.007$, $0.0075$, $0.008$, $0.0085$, $0009$, $0.0095$, $0.01$, $0.015$,
    $0.02$, $0.025$, $0.03\}$ \cite{Belczynski2020}.

Each metallicity bin requires a separate StarTrack simulation,
    but these metallicity bins are consistently organized
    across different time steps $\Delta t$.
In principle, it may be useful at this step to create a merger density
    $\rho_{\Lambda}(\BinaryParameters, t, \metallicity)$,
    by isolating $\rho_{\Lambda, \metallicity}(\BinaryParameters)$
    at each time step.
To conserve computation, we refrain from doing this.
Instead, we construct a weight function for
    the merger rate in a particular time step, for each metallicity bin:
\begin{equation}\label{eq:f_fr}
\rho_{\Lambda, \metallicity, \Delta t}(\BinaryParameters) = 
    \frac{M_{\Lambda,\metallicity,\Delta t}}{M_{\Lambda,\{\metallicity\}\Delta t}}
    \frac{\mathrm{SFR(\redshift)}}{M_{\mathrm{sim}}}
\end{equation}
Following this, a discrete marginalization yields the merger density
    for a synthetic universe:
\begin{equation}\label{eq:si}
\rho_{\Lambda}(\BinaryParameters) = \
    \sum\limits_{\Delta t, \metallicity}
        \rho_{\Lambda,\metallicity,\Delta t}(\BinaryParameters)
        \Delta t \Delta \metallicity
\end{equation}
This density function is scaled physically to represent
    merging binaries in 1 cubic megaparsec \correction{($\mathrm{Mpc}^3$)} in one year,
    with a correctly justified distribution in metallicity,
    and scaling in mass.
Hence, $\rho_{\Lambda}$ has units of \correction{
    mergers $\mathrm{Mpc}^{-3} \mathrm{yr}^{-1} $.
}
\end{subsection}
\begin{subsection}{Predicting the Gravitational-Wave Detection Rates}
\label{sec:Rdet}

While the merger rate, $\rho_{\Lambda}(\BinaryParameters)$ is useful
    in its own rite, we are primarily interested in those 
    simulated mergers which could be detected by current instrumentation.
We use a simplified detection model,
    considering observable simulated detections to require
    a Signal to Noise Ratio (SNR) of at least eight,
    using an interpolated SNR
    consistent with the LSC's predictions for instrument
    sensitivity (see \cite{P1200087}) 
    \correction{for a single detector network}.
We define a detection rate, $R(\BinaryParameters)$ for a sample
    merger, as
\begin{align}\label{eq:R_det}
R_{\Lambda}(\BinaryParameters) =
    \rho(\BinaryParameters)
    p_{\mathrm{det}}(\BinaryParameters)
    \frac{\mathrm{d}V_c}{\mathrm{d}\redshift_m}
\frac{\mathrm{d}t_m}{\mathrm{d}t_{det}}
\end{align}
where $\mathrm{d}t_m/\mathrm{d}t_{det} = 1/(1 + \redshift_m)$ is the factor relating
    merger time and detection time.
$p_{\mathrm{det}}$ is related to the LIGO sensitivity, and is interpolated
    from a table tabulated by other groups 
\cite{DominikIII}.
Finally, we also use
\begin{align}\label{eq:DominikIII-eq6}
\frac{\mathrm{d}V_c}{\mathrm{d}\redshift_m} =
\frac{4 \pi c}{H_0} \frac{D^2_c(\redshift)}{E(\redshift)}
\end{align}
where $D(\redshift)$ is the comoving distance and $E(\redshift)$ is a cosmological scale
    factor \cite{DominikIII}.

The expected number of gravitational-wave detections 
    for a given type of observation (BBH, BNS, NSBH)
    is calculated using that density according to Dominik et al.
    \cite{DominikIII}:
\begin{align}\label{eq:detection-rate}
\mu_{\Lambda, \alpha} = T_{\mathrm{obs}} \mathcal{R}_{\Lambda, \alpha}
    = \iiint\limits_{0}^{\infty}
    R_{\Lambda, \alpha}(m_1, m_2, \redshift_m) 
    \mathrm{d}\redshift_m \mathrm{d}m_1 \mathrm{d}m_2
\end{align}
where $R_{\Lambda, \alpha}(m_1,m_2,\redshift_m)$ 
    has been integrated
    over all merger parameters $\BinaryParameters$ except $m_1$, $m_2$, 
    and merger redshift for merger type $\alpha$.
$T_{\mathrm{obs}}$ is the time that at least two detectors were running
    during a particular gravitational-wave observing run.

\end{subsection}
\end{section}

\begin{section}{Simulation Properties}

\label{sec:population-properties}
\begin{table}
\centering
\begin{tabular}{|l|l|l|}
\hline
Series & Parameters Varied ($\Lambda$) & Supernova Engine \\
\hline \hline
M & Many Parameters Varied & Varied\\
\hline
M13-M19 & $\STkick$ & Rapid (strong pair-instability)\\
\hline
R & $\STFa$, $\STbeta$, $\STkick$ & Rapid (weak pair-instability)\\
\hline
D & $\STFa$, $\STbeta$, $\STkick$, $\STwind = \STwindHe$ & Delayed (weak pair-instability)\\
\hline
\end{tabular}
\caption{\label{tab:popsyn-families}
}
Families of StarTrack synthetic universes included in this dissertation.
Note that M-series models are the same as those that appear in 
    Belczynski et al. (2020) \cite{Belczynski2020}.
\end{table}

Now we discuss the properties of the StarTrack synthetic universes
    included in this study.
We consider several families of models (see table \ref{tab:popsyn-families}),
    with different assumptions about
    the formation parameters for binary evolution ($\Lambda$).
The primary formation parameters varied in each simulation include:
    supernova recoil kick velocity applied to 
    compact objects after a supernova explosion ($\STkick$),
    two parameters characterizing ejected mass and mass transfer
    ($\STFa$ and $\STbeta$),
    and the mass loss from common envelope evolution due to stellar wind,
    ($\STwind$).
We also consider rapid and delayed supernova engines with varying assumptions
    about pair-instability.
For each of these synthetic universes, we estimate the merger rate density
    $\rho_{\Lambda}(\BinaryParameters)$ and detection rate density 
    $\mathcal{R}_{\Lambda}(\BinaryParameters)$ for the entire population
    (see sections \ref{sec:s_i} and \ref{sec:Rdet}).
A full Bayesian comparison of the predicted gravitational-wave detection
    rate for each model to real observations is the focus of the next chapter.
In advance of this, we are interested in exploring 
    these properties of the merger population representing each model.

\begin{subsection}{Initial Models}
\label{sec:M-series-properties}
\begin{figure}
\centering
\includegraphics[width=0.6\columnwidth]{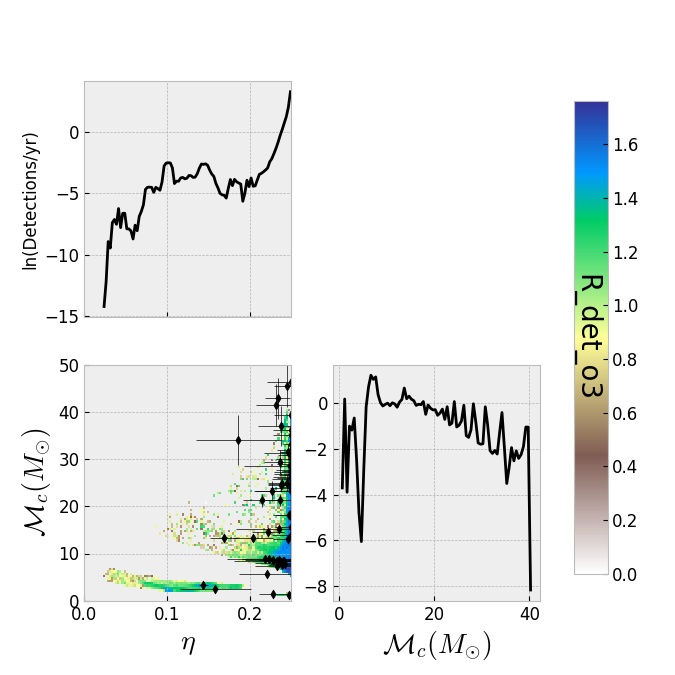}
\caption{\label{fig:M15}
Detection-rate weighted histograms for the sample mergers
    in the M15 StarTrack synthetic universe.
Diagonal elements indicate one-dimensional histograms
    (log base e),
    while the off-diagonal element shows
    a two-dimensional histogram (where color indicates
    the detection rate ($yr^{-1}$) for each bin; log base e).
MLE estimates for real gravitational-wave events
    included in our study are indicated by black diamonds,
    with error bars representing the characteristic $\sigma$
    associated with NAL models for each event.
}
\end{figure}

\begin{table}
\centering
\begin{tabular}{|l|l|}
\hline
Model & $\STkick (\mathrm{km/s})$ \\
\hline
M13 & 265 \\
\hline
M14 & 200 \\
\hline
M15 & 130 \\
\hline
M16 & 70\\
\hline
M17 & 35 \\
\hline
M18 & 20 \\
\hline
M19 & 10 \\
\hline
\end{tabular}
\caption{\label{tab:M13-19}
The $\STkick$ velocity is varied for the M13-M19 models.
}
\end{table}

The first set of models we study are those from Belczynski et al. (2016),
    which are labeled consistently with that study (M10, M15, etc...)
    \cite{Belczynski2016Nature}.
The only exception is M19, which hasn't been presented before.
We refer to these as the M-series models,
    and these models explore a large dimensional parameter space,
    effectively narrowing down our search for the ideal set of parameters.
Within the M-series, the M13-M19 models hold all formation parameters
    constant except for the supernova recoil kick velocity associated with
    BBH mergers, $\STkick$ 
    (section \ref{sec:st-formation-params}),
    which is varied between $10-265$ km/s \correction{(see table \ref{tab:M13-19})}.
This subset of the M-series models holds $\STFa=0.5$, $\STbeta=1.0$, 
    and $\STwind=1.0$.
These M-series models feature a rapid supernova engine with strong PPSN/PSN.
    \cite{Belczynski2020,Leung2019}.

We take a closer look at one of these models (M15) in particular,
    to describe the properties of a simulation before drawing an inference
    in the larger parameter space.
The cosmological merger rates represent a density in merger parameters
    $\rho_{\Lambda}(\BinaryParameters)$, for a population representing
    one set of formation parameters, $\Lambda$.
We see the detection rate as a function of source parameters
    ($\mathcal{M}_c$ and $\eta$).
For each merger in the simulation for M15,
    the detection rate is estimated as in eq. \ref{eq:R_det}.
Therefore, figure \ref{fig:M15} represents not simply the density of
    sample mergers in a synthetic universe predicted by M15,
    but the density of those mergers which would be detected
    by a gravitational-wave observatory in such a universe.

The M15 model is a good fit to the observed detection rate
\correction{
    (see fig \ref{fig:M-series-detections}; left panel).
}
The substantial mass gap present for this model
    is indicative of the rapid supernova engine.
This model is able to produce most events observed by the collaboration,
    but fails to desribe mass gap events
    and under-predicts high mass events.

\begin{figure}
\centering
\includegraphics[width=0.48\columnwidth]{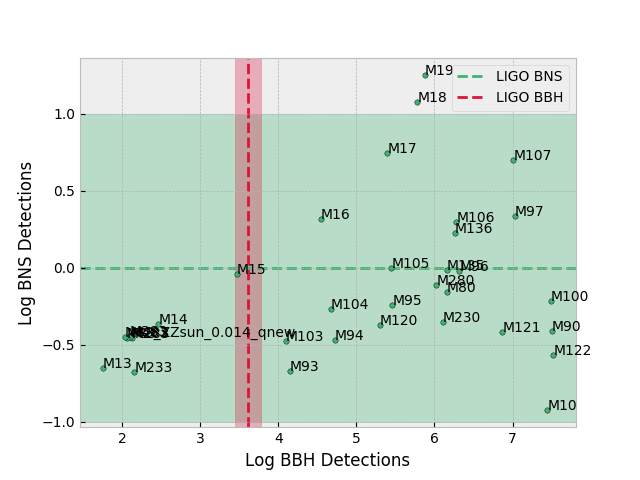}
\includegraphics[width=0.48\columnwidth]{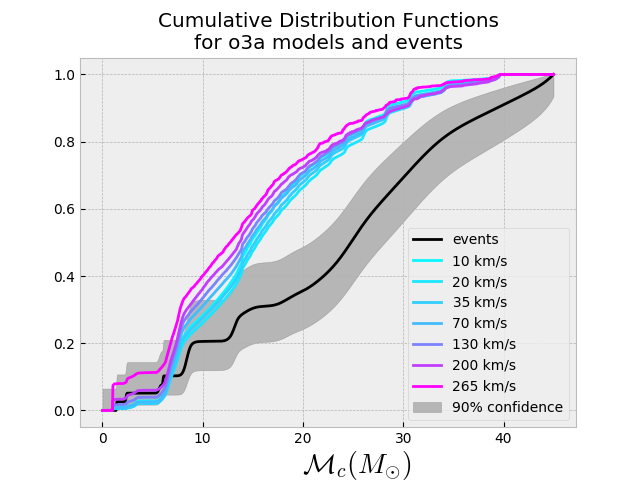}
\caption{\label{fig:M-series-detections}
Detection rates for the limited M-series are compared to the first part of the
    third observing run (GWTC-2).
(left)
The detection rates predicted (log base e) 
    for binary black holes and neutron stars
    are labeled for individual StarTrack models 
    in this scatter plot.
The dotted lines represent the number of observed detections for a given kind
    in GWTC-2,
    and shaded regions indicate Poisson counting error.
(right)
CDFs in source-frame chirp mass $\mathcal{M}_c$
    (similar to figure \ref{fig:nal-cdf}).
Thin colored lines demonstrate the 
    predicted gravitational-wave events in the M-series population models,
    (colored by $\STkick$).
\correction{
The shaded region represents a 90\% symmetric credible Wilson score interval
    \cite{WilsonScore}.
}
}
\end{figure}

Trends in the observed detection rate for M-series models
    can be seen in figure \ref{fig:M-series-detections}.
The scatter plot plot compares the predicted binary black hole
    event rates and binary neutron star event rates for population
    models in the M series.
The CDF plot compares the shape of each detection-weighted
    sample density in each StarTrack synthetic universe.
In the next chapter
    (sec \ref{sec:M-series-likelihood}),
    we will examine how well the properties of these
    detection rate predictions agree with the observations
    reported in the GWTCs.

\end{subsection}

\begin{subsection}{Preliminary Uniform Parameter Space}
\label{sec:R-series-properties}

\begin{figure}
\centering
\includegraphics[width=0.48\columnwidth]{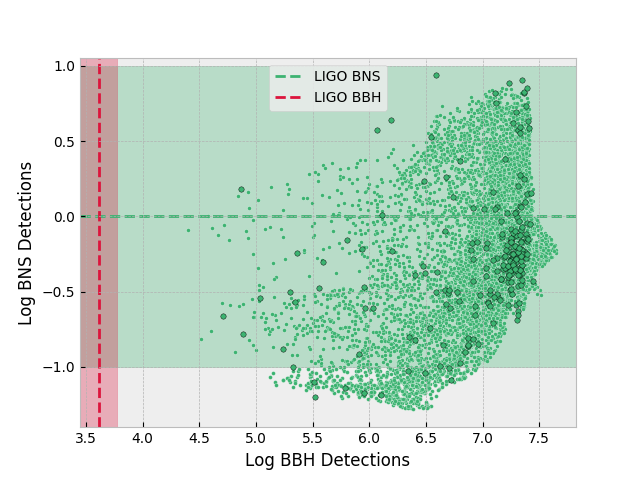}
\includegraphics[width=0.48\columnwidth]{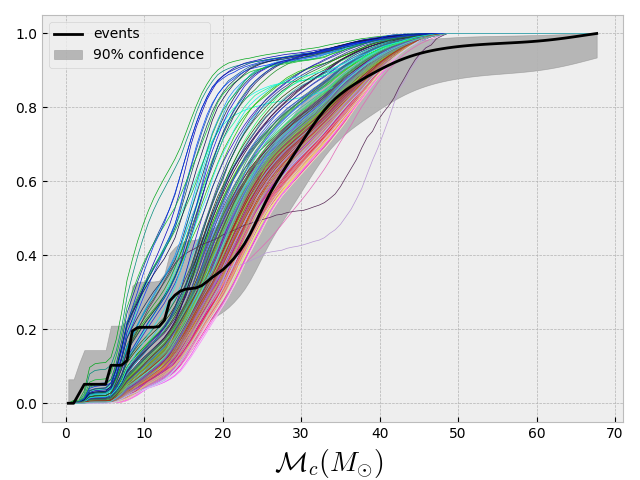}
\caption{\label{fig:R-series-detections}
Detection rates for the R-series are compared to the first part of the
    third observing run (GWTC-2).
Structurally similar to figure \ref{fig:M-series-detections}.
(left)
The smaller points are drawn from interpolated models of the larger
    parameter space of the R series simulations.
(right)
The color scale for R-series model CDFs encode the three formation parameters
    as Red, Green, and Blue values for $\STFa$, $\STbeta$, and $\STkick$.
\correction{
For the R-series models,
    we study a wider mass range than before,
    with the goal of reproducing the more massive BBH mergers observed
    in the third observing run.
}
}
\end{figure}

Following the one-dimensional study,
    the R-series models explore a formation parameter space given by
    300 randomly sampled points in the space of
    $\STFa$, $\STbeta$, $\STkick$,
    with a rapid supernova engine,
    fixed $\STwind=0.2$,
    and weak pair-instability.
$\STFa$ is varied between 0 and 1.
$\STbeta$ is varied between 0 and 1.
$\STkick$ is varied between 0 and $265$ km/s.
The reduced mass loss due to stellar wind was expected
    to assist in producing high mass events such as 
    the observed event, GW190521,
    through isolated binary evolution alone.
However, as seen in figure \ref{fig:R-series-detections},
    the R-series models overpredict gravitational-wave 
    detection rates by several orders of magnitude.
These detection rates lead us to believe that some of our assumptions in
    the formation parameters of these models may have been wrong.
This is explored in greater detail in section \ref{sec:D-series-likelihood}.

\end{subsection}
\begin{subsection}{Delayed Supernova Engine and Wind}
\label{sec:D-series-properties}

\begin{figure}
\centering
\includegraphics[width=0.48\columnwidth]{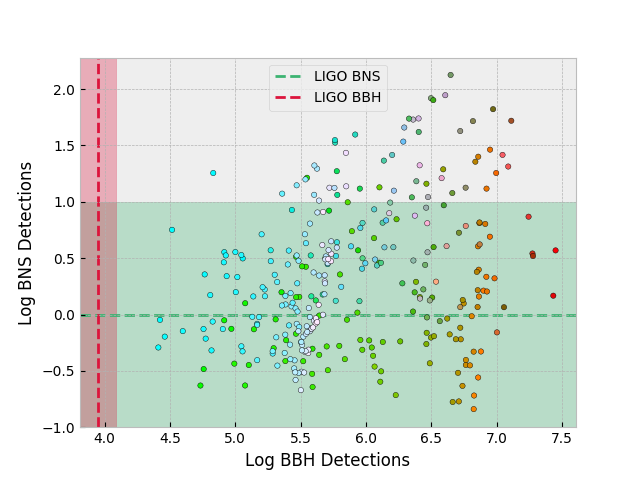}
\includegraphics[width=0.48\columnwidth]{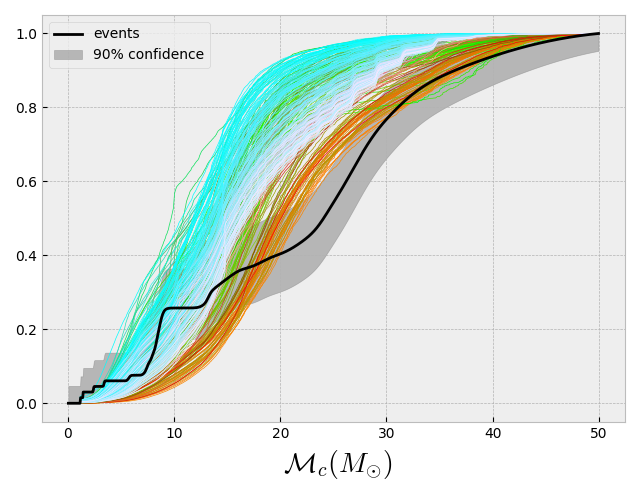}
\caption{\label{fig:D-series-detections}
Structurally similar to figure \ref{fig:M-series-detections}.
(left)
Detection rates for the D-series are compared to the third 
    observing run of (GWTC-2 and GWTC-3).
(right)
Detection rate CDFs for the D-series models are shown
    for comparison with all events for the first three observing runs
    (colored by an RGB scale in $\STFa$, the detection rate log likelihood;
    e.g. eq \ref{eq:rate-likelihood}, and $\STwind$).
\correction{
For the D-series mergers, our mass range highlights
    most of the high-mass systems observed in LIGO/Virgo's
    third observing run, however our binary evolution simulations
    do not result in any mergers like GW190521.
}}
\end{figure}

\begin{figure}
\centering
\includegraphics[width=0.48\columnwidth]{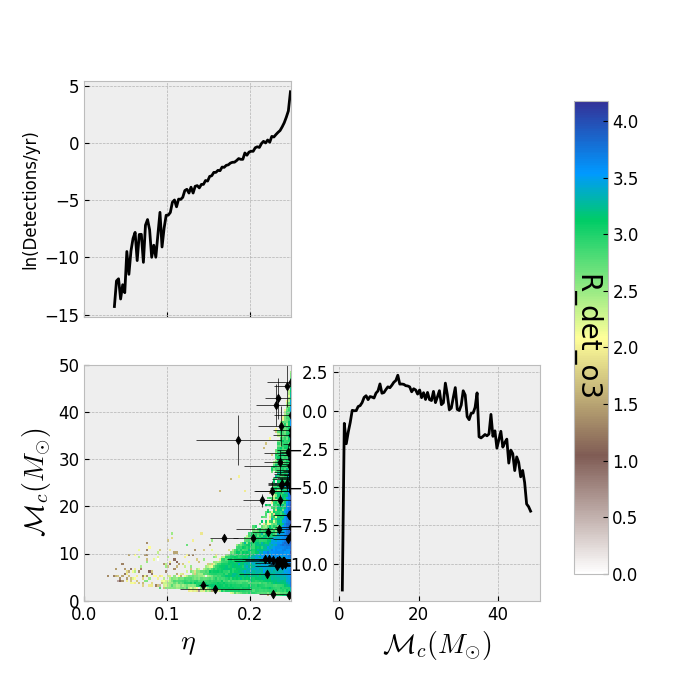}
\caption{\label{fig:D411}
Detection-rate weighted histograms for the sample mergers
    in the and \correction{D411} 
    StarTrack synthetic universe (log base e),
    in the style of figure \ref{fig:M15}.
}
\end{figure}

Following our studies in the one- and three- formation parameter
    spaces of the M- and R-series synthetic universes,
    the properties of those models convinced us (see section \ref{sec:D-series-likelihood})
    to explore the formation parameter $\STwind$ as well,
    with an updated delayed supernova engine and weak pair-instability,
    consistent with Belczynski et al. (2020) \cite{Belczynski2020,Leung2019}.
The formation parameter space for the D-series synthetic universe
    models
    is uniformly distributed in $\STFa$, $\STbeta$, and $\STkick$
    like the R-series models \correction{(but with $0 < \STFa < 0.25$).}
Of the 300 models generated for this series,
    100 models are held at fixed $\STwind=0.2$,
    100 models at fixed $\STwind=1.0$, 
    and 100 models varied uniformly in $\STwind$ between those values.
We note that the detection rate found for these models is more
    consistent with observations than the R-series
    models,
    and a full likelihood analysis follows in section
    \ref{sec:D-series-likelihood}.

We include figure \ref{fig:D411},
    which shows the difference in the predicted population of
    gravitational-wave detections with a delayed supernova engine.
Notice the lack of a substantial mass gap, compared to M15
    (fig \ref{fig:M15}).
For the delayed engine models, 
    \correction{
    the D411 synthetic universe is our best match to observations
    (see chapter \ref{chap:popsyn}),
    with $\STFa=0.0185$, $\STbeta=0.94$, $\STkick=205$km/s,
    and $\STwind=1.0$.}

\end{subsection}
\end{section}
 \end{chapter}

\begin{chapter}{Population Synthesis}
\label{chap:popsyn}
The previous chapter demonstrated our procedure for    
    estimating sample merger densities for massive binaries
    in synthetic universes generated with the StarTrack population synthesis
    models for a choice of formation parameters
    $\rho_{\Lambda}(\BinaryParameters)$.
We have also explored the properties of these merger populations,
    including the predicted gravitational-wave detection rates.
Now we explore the Bayesian population inference framework that allows us
    to constrain these binary evolution processes in a space of
    these formation parameters.
This analysis is built upon the Bayesian framework described by other groups
    such as Dominik et al. and Wysocki et al.
    \cite{DominikIII, Wysocki2018, Wysocki2019}.
We study a one-, three-, and four-dimensional parameter space of
    StarTrack population synthesis models,
    and use these simulations to constrain the evolution of massive
    stellar binaries.
These results draw heavily upon my forthcoming publication which 
    has been presented internally to the LSC in 2020
    and presented at the American Physical Society's April meeting in 2021.

\begin{section}{Bayesian Population Inference}
\label{sec:popsyn-likelihood}

The observation of gravitational-wave events can be modeled as
    an Inhomogeneous Poisson Point Process
    \cite{Wysocki2019}.
This representation allows us to evaluate the probability
    of the set of observed detections in
    the GWTC releases, $\gwdets$,
    for a simulated merger population, 
    $\rho_{\Lambda}(\BinaryParameters)$.
We must evaluate the relative likelihood that a population model 
    accurately describes the detections observed by
    gravitational-wave detectors.
This quantity, $\pprob(\gwdets|\Lambda)$,
    equivalent to calculations done by others in the field
    \cite{Wysocki2019, LIGO-O3-O3a-RP},
    can be decomposed as such:
\begin{align}
    \pprob(\gwdets|\Lambda) = 
        \pprob(\mu|\Lambda) \prod\limits_j \pprob(\gwdetj|\Lambda)
\end{align}
where the agreement of the predicted detection rate, $\mu$,
    for a synthetic universe described by the formation parameters,
    $\Lambda$, with the observed detection rate in the GWTC releases
    for each type of event (E.g. BBH, BNS, NSBH) are described by
    simple Poisson Point Process components, and contained in
    the ``rate likelihood'' $\pprob(\mu|\Lambda)$.
Meanwhile, the agreement of the properties of sample mergers 
    in that synthetic universe with the likelihood function 
    attributed to the parameters of each
    real gravitational-wave detection
    is contained in ``shape likelihood''
    $\pprob(\gwdetj|\Lambda)$.

\correction{
We make use of Gaussian Process regression
    in order to interpolate these results,
    so as to describe the parameter space between simulations.
The interpolation is carried out using the gp-api
    software package developed by our group
    (see section \ref{sec:gp-api}),
    using a kernel with piecewise polynomial basis functions
    with compact support, as outlined in Williams and Rasmussen
    \cite{williams2006gaussian}.
The scale coefficient hyperparameters for the kernel are chosen
    by hand for the interpolations carried out in this chapter.
}

\begin{subsection}{Detection Rate Likelihood}
We have previously calculated these detection rates,
    $\mathcal{R}_{\Lambda, \alpha}$,
    and predicted number of detections, $\mu_{\Lambda,\alpha}$,
    for each
    of the population models described in 
    section \ref{sec:population-properties}
    (see eq. \ref{eq:detection-rate}).
The observation time, $T_{\mathrm{obs}}$, assumed is consistent
    with GWTC release estimates for the time at least 2 detectors
    were running.
The probability of an observed number of detections
    for a given type of event ($\alpha \in \{\mathrm{BBH, BNS, NSBH}\}$),
    is dependent on the predicted rate
    for each type of event, described by the Poisson Point Process:
\begin{align}\label{eq:poisson-likelihood}
\pprob(\mu_{\alpha} | \Lambda) = 
    e^{-(\mathcal{R}_{\Lambda, \alpha} T_{\mathrm{obs}})} 
    \frac{(\mathcal{R}_{\Lambda, \alpha} T_{\mathrm{obs}})^{N_\alpha}}{N_{\alpha}!}
\end{align}
where $N_{\alpha}$ is the number of gravitational-wave detections
    of type $\alpha$ reported in the GWTCs.
Alternatively, the log rate likelihood is
\begin{align}\label{eq:poisson-log}
\mathrm{ln}(\pprob(\mu_{\alpha} | \Lambda)) =
    -(\mathcal{R}_{\Lambda, \alpha} T_{\mathrm{obs}})
    + N_{\alpha}\mathrm{ln}(\mathcal{R}_{\Lambda,\alpha})
    + N_{\alpha}\mathrm{ln}(T_{\mathrm{obs}})
    - \mathrm{ln}(N_{\alpha}!)
\end{align}
where the last two terms do not depend on the properties of
    the synthetic universe, $\Lambda$, and can be ignored
    to a factor of renormalization.

The total combined rate likelihood is therefore
\begin{align}\label{eq:rate-likelihood}
\pprob(\mu | \Lambda) = 
    \prod\limits_{\alpha} \Big[
        \frac{(\mathcal{R}_{\Lambda,\alpha}T_{\mathrm{obs}})^{N_{\alpha}}}{N_{\alpha}!}
        \Big]
        e^{-\sum\limits_{\alpha} \mu_{\Lambda,\alpha}}
        = \mathcal{K}_{\mathrm{rate}} e^{-\mu_{\Lambda}}
\end{align}
Or alternatively,
\begin{align}\label{eq:rate-likelihood-log}
\mathrm{ln}(\pprob(\mu | \Lambda)) \propto
    \sum\limits_{\alpha} 
    (- \mathcal{R}_{\Lambda,\alpha}T_{\mathrm{obs}} +
    N_{\alpha}\mathrm{ln}(\mathcal{R}_{\Lambda,\alpha}))
    = -\mu_{\Lambda} + \sum\limits_{\alpha}
    N_{\alpha}\mathrm{ln}(\mathcal{R}_{\Lambda,\alpha})
\end{align}
\end{subsection}

\begin{subsection}{Shape Likelihood}

The agreement of the sample mergers in a StarTrack
    synthetic universe with the properties of the gravitational-wave
    signals observed thus far by gravitational-wave detectors
    is measured by the Inhomogeneous Poisson likelihood,
    as stated above.
The ``shape likelihood'' is the contribution toward this likelihood
    from the shape of the distribution of observed detections.
This quantity measures the agreement of the sample merger density,
    $\rho_{\Lambda}(\BinaryParameters)$,
    for a synthetic universe (represented by some formation parameters,
    $\Lambda$)
    with the likelihood for the astrophysical parameters of each
    gravitational-wave detection,
    $\mathcal{L}_j(\BinaryParameters)$.
This quantity is marginalized (integrated) over
    the entire population of sample mergers in a synthetic universe:
\begin{align}\label{eq:shape-likelihood}
\pprob(d_j|\Lambda) =
    \int\limits_{\BinaryParameters} 
        \pprob(d_j|\BinaryParameters,\Lambda) P(\BinaryParameters|\Lambda)
        \mathrm{d}\BinaryParameters =
    \int\limits_{\BinaryParameters}
        \bar{\rho}_{\Lambda}(\BinaryParameters) \mathcal{L}_j(\BinaryParameters)
        \mathrm{d}\BinaryParameters
\end{align}
where $\bar{\rho_{\Lambda}}(\BinaryParameters) = \rho_{\Lambda}(\BinaryParameters)
    / \int_{\BinaryParameters'} \rho_{\Lambda}(\BinaryParameters') d\BinaryParameters$.

The marginalization is carried out over the entire population of
    $~10^8$ sample compact binary mergers in a each synthetic universe,
    for each gravitational-wave observation from a given observing run.
In this chapter, we consider the likelihood contribution from
    the same events included in the LSC's O3 rates and populations
    paper
    \cite{LIGO-O3-O3b-RP},
    with the exclusion of GW190521, which we expect
    may be inconsistent with isolated binary evolution
    (see sec \ref{sec:D-series-likelihood}).
The discrete evaluation would not be computationally reasonable
    without an approximate model for each likelihood
    that can evaluated in a small fraction of second.
We make use of the NAL estimates provided in chapter \ref{chap:nal}
    for this integration.

\end{subsection}

\begin{subsection}{Joint Likelihood}
Our complete expression for the joint likelihood
    agrees with Wysocki et al. 2019 \cite{Wysocki2019}:
\begin{align}\label{eq:wysocki-likelihood}
\pprob(\gwdets|\Lambda) \propto
    K_{\mathrm{rate},i} \mathrm{e}^{-\mu}
    \prod\limits_j \Big[
        \int\limits_{\BinaryParameters}
        \bar{\rho}(\BinaryParameters) \mathcal{L}_j(\BinaryParameters)
        \mathrm{d}\BinaryParameters
        \Big]
\end{align}
Alternatively the logarithmic representation:
\begin{align}\label{eq:wysocki-log}
\mathrm{ln}(\pprob(\gwdets|\Lambda)) \propto
    -\mu_{\Lambda} + \sum\limits_{\alpha}[N_{\alpha} \mathcal{R}_{\Lambda,\alpha}]
        + \sum\limits_{j} \Big[ \int\limits_{\BinaryParameters}
            \bar{\rho}(\BinaryParameters) \mathcal{L}_j (\BinaryParameters)
            \mathrm{d}\BinaryParameters\Big]
\end{align}
\end{subsection}
\end{section}

\begin{section}{Kick Velocity; A One-Dimensional Study}
\label{sec:M-series-likelihood}

\begin{figure}
\centering
\includegraphics[width=0.48\columnwidth]{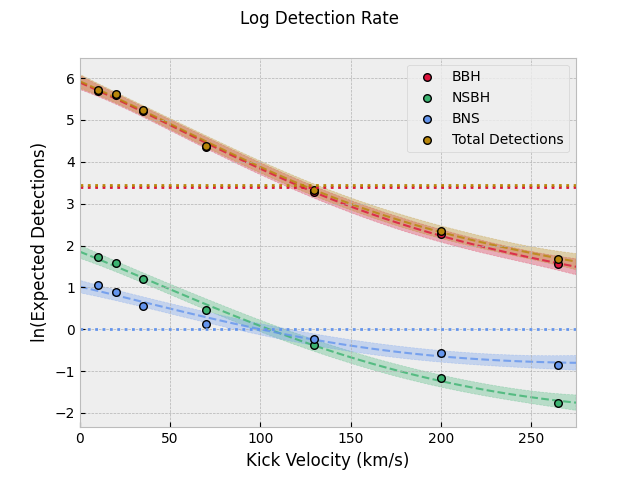}
\includegraphics[width=0.48\columnwidth]{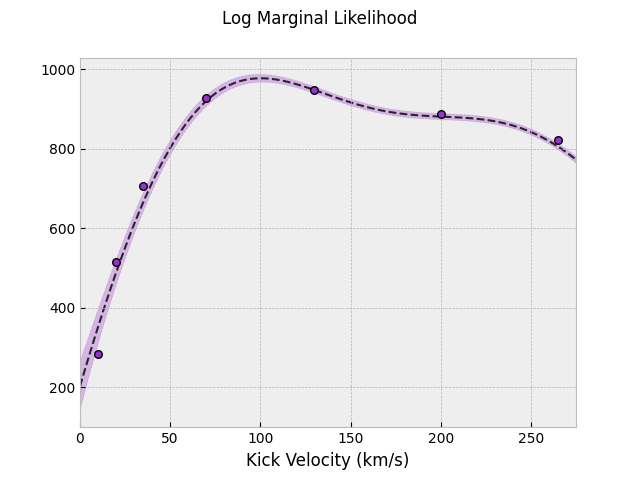}
\caption{\label{fig:M-interp}
Detection rates (left) and the joint likelihood (right)
    for various gravitational-wave events
    as function of the kick velocity formation parameter.
Scatter points indicate model simulations (M13-M19).
The dashed curve indicates the mean from
    a Gaussian process regression interpolation.
\correction{For this example only,
    scikit-learn is used for the interpolation due
    to the small number of training points}.
The shaded region is of one standard deviation error from
    that Gaussian process.
The dotted line (left) indicates the published number of detections
    for the first part of the LIGO's third observing run.
There is a clear peak in likelihood near M15, where the kick velocity 
    is 130 km/s, and where the LIGO detection rates also agree.
}
\end{figure}

The M-series models shown in figure \ref{fig:M-series-detections}
    include many families of cosmological formation parameters $\Lambda$.
In our first one-dimensional study,
    we limit ourselves to the models M13-M19, 
    which hold everything constant except $\STkick$
    which varies from $10$ km/s to $265$ km/s
    (see sec \ref{sec:M-series-properties}).
We note that the contribution from individual events 
    (shape likelihood) dominates
    the joint likelihood for this family of models,
    changing more rapidly between
    populations than the rate likelihood.
As seen in figures \ref{fig:M-series-detections} and \ref{fig:M-interp},
    we see the best agreement with model M15,
    with $\sigma = 130 \mathrm{km/s}$.
    and our analysis are in favor of modest supernova recoil kicks
    for black holes.

In the space of the kick velocity formation parameter,
    we have generated an interpolation 
    using a Gaussian Process Regression algorithm.
The Gaussian process helps to describe
    the detection rate for each type of event
    in the space between simulations.
It also describes the joint likelihood in that space.
We find that a modest kick velocity is appropriate for 
    our model of binary evolution.

\end{section}
\begin{section}{Higher-Dimensional Studies: $\STFa$, $\STbeta$, $\STkick$, $\STwind$}
\label{sec:D-series-likelihood}

Following the one-dimensional investigation,
    the next set of models we study is a three-dimensional
    formation parameter space,
    constraining $\STFa$, $\STbeta$, and $\STkick$
    using an updated rapid supernova engine.
As seen in section \ref{sec:R-series-properties}
    (figure \ref{fig:R-series-detections}),
    the R-series models do not predict similar detection rates
    to the one-dimensional M-series study
    and are inconsistent with observation.
This can be explained by differences in the supernova engine,
    together with differences in the mass loss due to stellar wind,
    $\STwind$.
Unfortunately, the decreased mass loss due to stellar winds
    assumed in the R-series models predicts an overabundance of 
    binary black hole detections for this set of models
    by several orders of magnitude.
For these models,
    the inconsistency with the observed gravitational-wave
    detection rate dominates likelihood calculation
    (the rate likelihood contribution; 
    see fig \ref{fig:supernova-detection}).

In hindsight, the choice of $\STwind=0.2$ for these models
    proved to be an optimistic attempt to produce events such as
    GW190521 by isolated binary evolution.
Recent studies have argued GW190521 has properties
    (such as orbital eccentricity at time of merger)
    which are not consistent with isolated binary evolution,
    and may be the result of a dynamic (3 or more bodies) merger
    \cite{Gayathri2022, Gamba2021, Romero_Shaw_2020}.
In light of this,
    and because it is unreasonable to expect
    that all of the marginal detections reported in GWTC-2.1
    are of astrophysical origin \cite{GWTC-2p1},
    we must not demand that our StarTrack synthetic universe can
    predict every single merger reported in the GWTC releases.
Doing so may over-constrain the parameters of the isolated
    binary evolution channel to produce unrealistic results.

Finally, we have also generated a
    tertiary set of merger populations varying the same three parameters,
    using a delayed supernova engine, and also varying 
    the factor of reduced mass loss, $\STwind$ assumed by contemporary 
    literature \cite{Belczynski2020}.
The D-series models provide more compelling constraints on the astrophysical
    assumptions about binary evolution encoded in these parameters.

\begin{subsection}{The Delayed Supernova Engine}

\begin{figure}
\centering
\includegraphics[width=0.49\columnwidth]{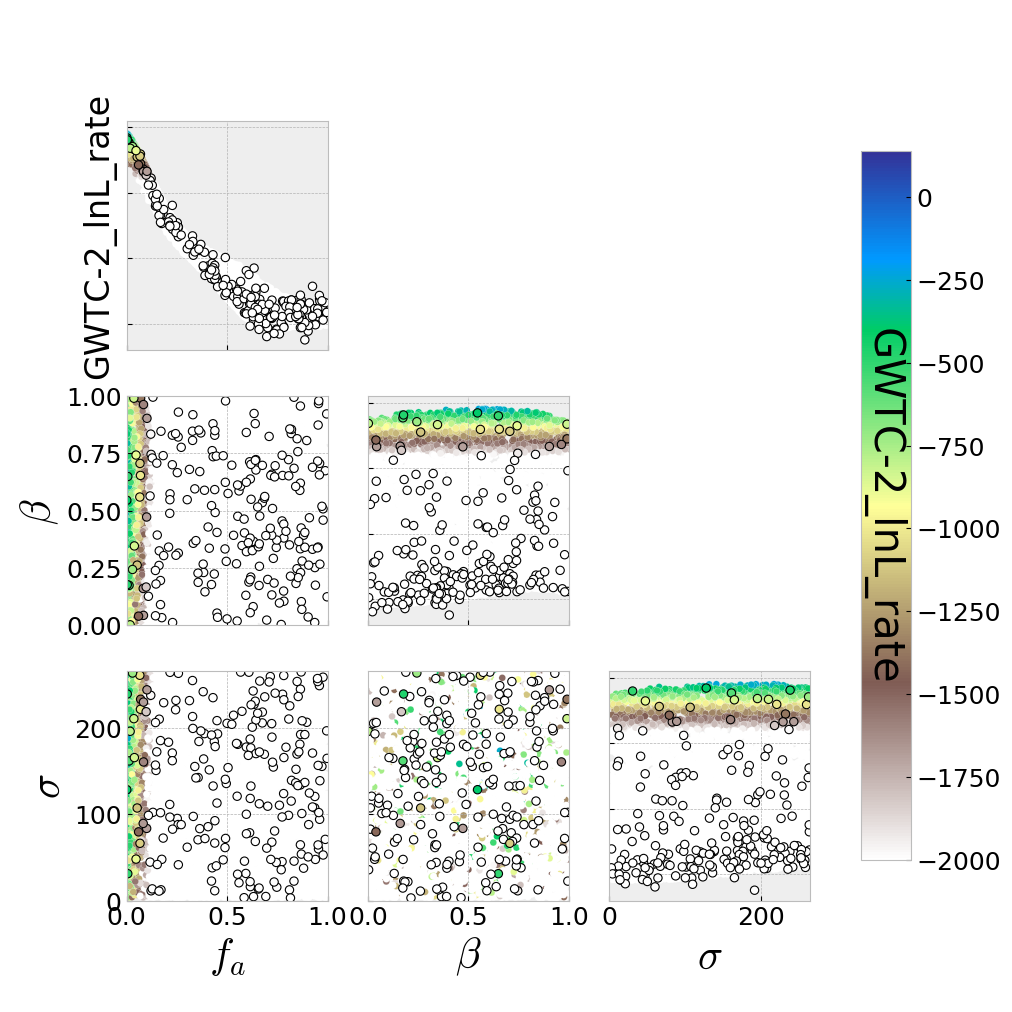}
\includegraphics[width=0.49\columnwidth]{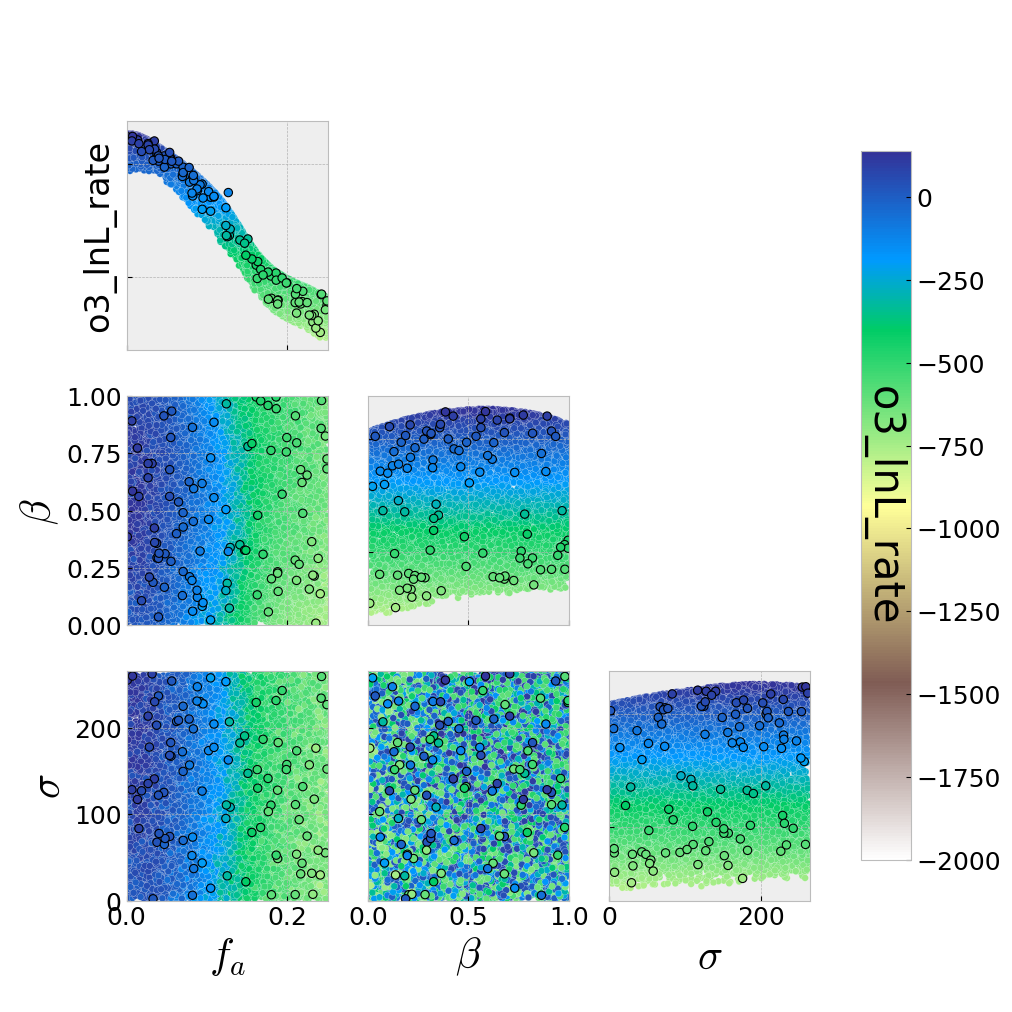}
\caption{\label{fig:supernova-detection}
}
Scatter plots highlight the relative probability with which synthetic universes
    with formation parameters varied in $\STFa$, $\STbeta$, and $\STkick$
    represent a realistic model,
    based on soley the detection rate likelihood in log scale (colorbar)
    (eq \ref{eq:rate-likelihood}, \ref{eq:rate-likelihood-log}).
Each synthetic universe for the R-series models (left), 
    and the subset of D-series models with $\STwind=0.2$ (right)
    are represented by a point with a black border.
A Gaussian process regression interpolation is used to estimate the rate 
    likelihood between synthetic universe models
    (smaller points with white borders).
Note the difference in scale in $\STFa$.
We consider the first part of the third observing run for R-series models,  
    and the entire third observing run for D-series models.
We observe that synthetic universes with both supernova engines over-produce
    observable BBH events,
    which necessarily constrains $\STFa$ to be small,
    due to the high correlation between $\STFa$ and detection rates
    at low $\STwind$.
However, this overabundance is scaled down with
    the delayed engine supernova model.
\end{figure}

Before exploring the four-dimensional parameter space composing
    the D-series models,
    we take a moment to consider the same three-dimensional
    parameter space explored previously by the R-series.
The first 100 models in the D-series keep a fixed value of 
    $\STwind =0.2$, consistent with the R-series models.
Therefore, it is worth drawing a comparison between the R-series models
    and this reduced set of D-series models in order to characterize the
    effect of changing the supernova engine
    (see fig \ref{fig:supernova-detection}).

For both sets of models,
    we find a high correlation between the estimated detection rate
    and our $\STFa$ formation parameter,
    however this effect is less prominent for the delayed supernova engine.
Rather than drawing strong conclusions about the $\STFa$ parameter
    based on this correlation,
    we note that even for low $\STFa$, we still predict
    an overabundance of BBH detections
    (see fig \ref{fig:R-series-detections} and \ref{fig:D-series-detections}).
It is worth considering what kind of biases we may be introducing to our
    estimation of the predicted detection rates due to our detection model.

We use a Gaussian process between models to predict the detection rate
    likelihood in the entire parameter space studied in each series of models
    (\ref{sec:R-series-properties}, \ref{sec:D-series-properties}).
This method allows us to study a larger formation parameter space
    without an ever-increasing number of simulations.
We use our own Gaussian process regression aglorithm for this
    application (sec \ref{sec:gp-api}).

\end{subsection}
\begin{subsection}{Constraints on Isolated Binary Evolution in Four Dimensions}
\label{sec:constraints-4D}

\begin{figure}
\centering
\includegraphics[width=0.48\columnwidth]{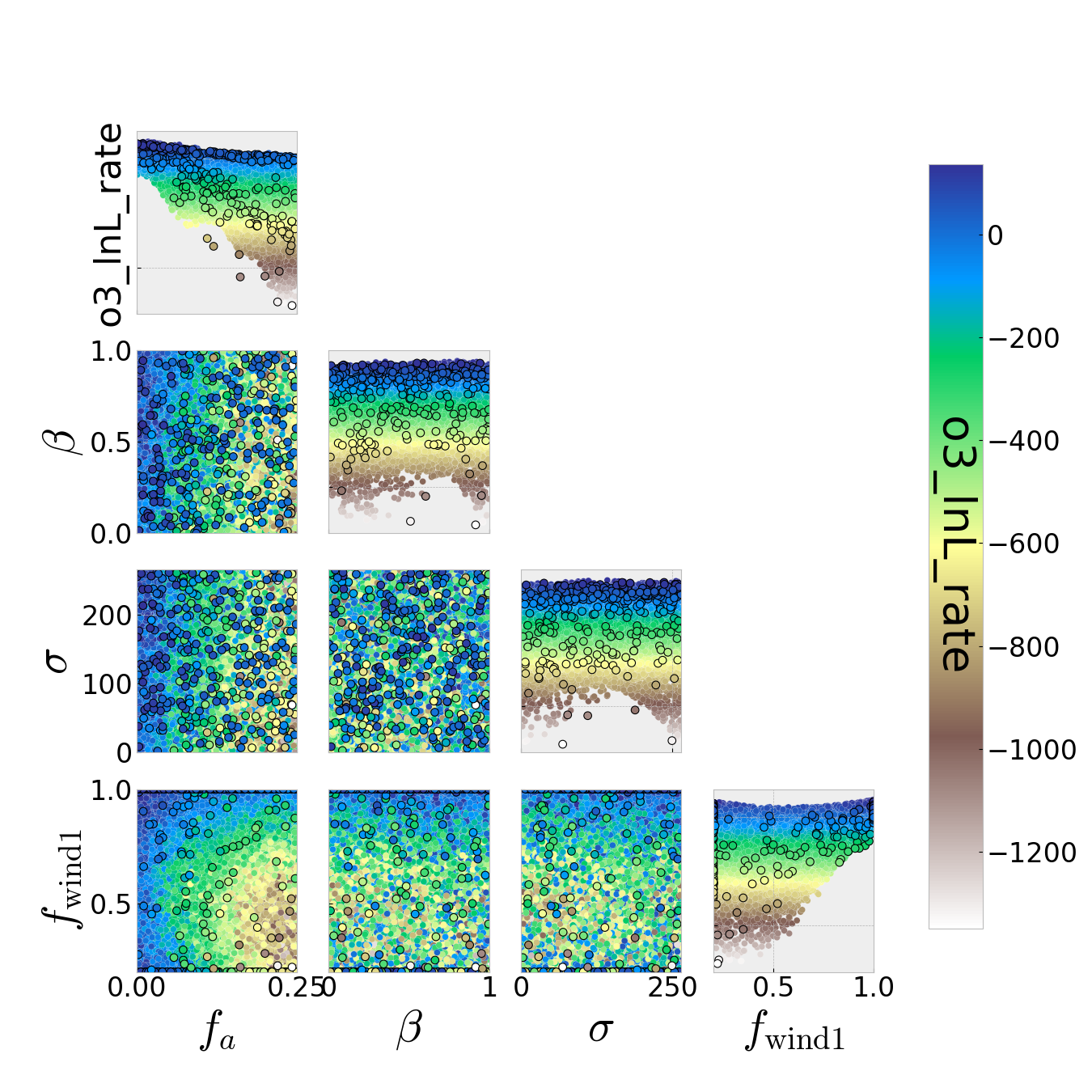}
\includegraphics[width=0.48\columnwidth]{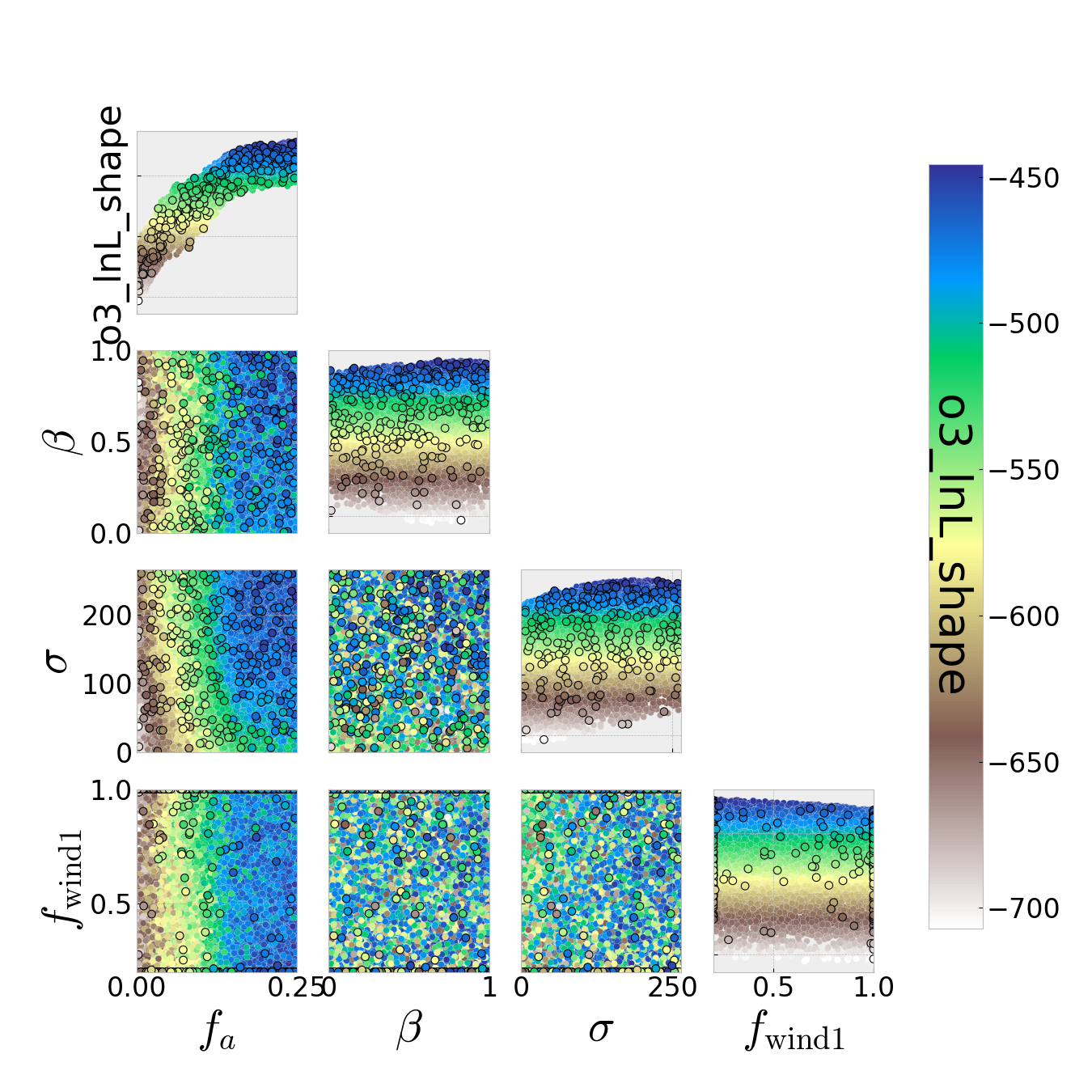}
\caption{\label{fig:D-series-likelihood}
The shape likelihood (left) and rate likelihood (right) for the full
    set of D-series synthetic universe models,
    using the same style as fig \ref{fig:supernova-detection}
    (log base e;
    see eq \ref{eq:rate-likelihood-log}, \ref{eq:shape-likelihood}).
Note that at high $\STwind$, the correlation between rate likelihood
    and $\STFa$ is reduced.
Note also that the shape likelihood favors $\STFa > 0$.
}
\end{figure}

\begin{figure}
\centering
\includegraphics[width=0.48\columnwidth]{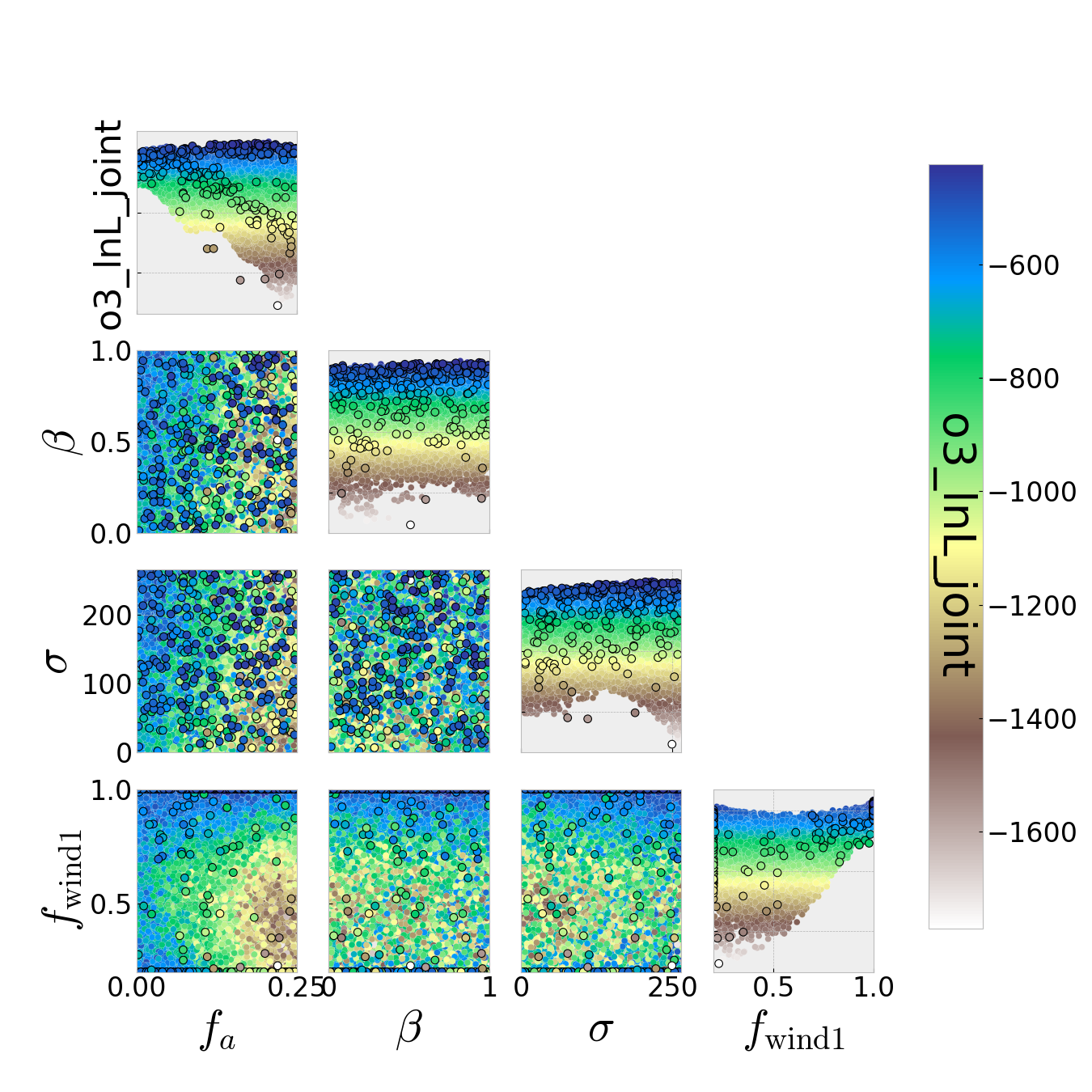}
\includegraphics[width=0.48\columnwidth]{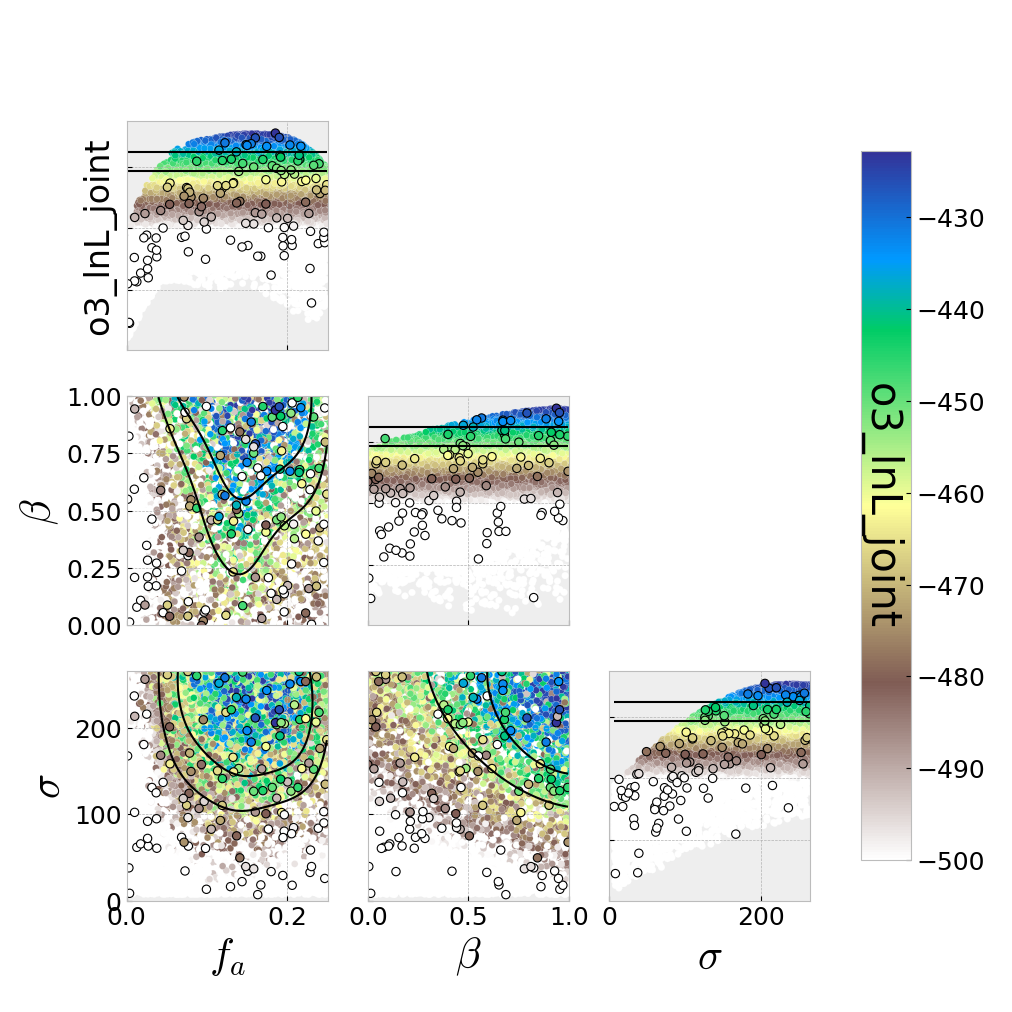}
\caption{\label{fig:D-series-joint-likelihood}
The joint likelihood (log base e; see eq \ref{eq:wysocki-log})
    for models in the D-series,
    in the same style as fig \ref{fig:supernova-detection}.
(left) the full set of models in four dimensions.
(right) a three-dimensional interpolation of only high wind models
    with horizontal lines and contours for regions of 1 and 2 $\sigma$ 
    in the Gaussian process estimated uncertainty.
}
\end{figure}
At last, we can use the full set of delayed supernova engine models
    to constrain the formation parameters of our StarTrack synthetic universe
    formula (see sec \ref{sec:D-series-properties}).
We show the (log base e) likelihood for each synthetic universe in figures
\ref{fig:D-series-likelihood} and \ref{fig:D-series-joint-likelihood},
    featuring scatter-plots and Gaussian process interpolations
    for the rate likelihood, shape likelihood, and joint likelihood
    (sec \ref{sec:popsyn-likelihood}).
We explore the implications of different choices for the
    formation parameters $\STFa$, $\STbeta$, $\STkick$, and $\STwind$
    on these quantities,
    and in doing so demonstrate our method for constraining 
    these processes in the isolated binary evolution formation
    channel for gravitational-wave events.

\begin{table}
\centering
\begin{tabular}{|l|l|l|l|l|l|l|l|}
\hline
Model & $\mathrm{ln}\mathcal{L}_{\mathrm{joint}}$ & $\mathrm{ln}\mathcal{L}_{\mathrm{rate}}$ & $\mathrm{ln}\mathcal{L}_{\mathrm{shape}}$ & $\STFa$ & $\STbeta$ & $\STkick$ (km/s) & $\STwind$ \\
\hline
D411 & 0.000 & -98.149 & -16.107 & 0.185 & 0.938 & 205.624 & 1.00 \\
D473 & -3.539 & -98.715 & -19.079 & 0.190 & 0.954 & 220.655 & 1.00 \\
D301 & -3.906 & -96.591 & -21.571 & 0.160 & 0.792 & 211.110 & 1.00 \\
D407 & -7.924 & -75.311 & -46.868 & 0.122 & 0.566 & 257.617 & 1.00 \\
D465 & -8.619 & -97.120 & -25.755 & 0.155 & 0.887 & 183.200 & 1.00 \\
\hline
D426 & -101.096 & 0.000 & -215.351 & 0.010 & 0.594 & 236.519 & 1.00 \\
\hline
D027 & -628.026 & -742.282 & 0.000 & 0.249 & 0.683 & 226.504 & 0.20 \\
\hline
\end{tabular}
\caption{\label{tab:D-series-stats}
\correction{
Quantities of interest for the simulations with the five best
    joint likelihood estimates, the best rate likelihood estimate,
    and the best shape likelihood estimate.
Tabulated values are re-scaled by the 
    by the highest likelihood for each category 
    ($\mathrm{ln}\mathcal{L}_{\mathrm{joint-max}} = -422.824$,
    $\mathrm{ln}\mathcal{L}_{\mathrm{rate-max}} = 136.966$,
    $\mathrm{ln}\mathcal{L}_{\mathrm{shape-max}} = -445.534$).
}}
\end{table}

Our method allows us to assess the quality of fit
    between each synthetic universe and the observed gravitational-wave
    population
    by separately evaluating the agreement
    in detection rate and the shape of the population.
This gives us the ability to study the dependence of
    the rate and shape likelihood components on the formation model
    independently.
When the rate and shape likelihood disagree,
    this gives us cause to re-examine our formation model.
When these quantities are in agreement
    the joint likelihood will be maximized,
    indicating the best potential candidate for a realistic
    set of formation model assumptions.

While the joint likelihood for the M-series models was strongly constrained by 
    the shape of the population,
    and the joint likelihood of the R-series models
    was dominated by the effect of the detection rate,
    we observe that with the delayed supernova engine,
    the rate likelihood and shape likelihood contributions
    are at similar orders of magnitude,
    and both contribute to the joint likelihood.
We observe a relative agreement from these models in some
    formation parameters.
The rate and shape likelihoods both support substantial supernova kicks,
    consistent with our M-series models.
We see also a tendency towards high $\STbeta$ in both likelihood
    estimates.

We also observe a strong codependence of the predicted detection
    rates on both mass loss rate due to stellar wind ($\STwind$)
    and $\STFa$.
However, the rate and shape likelihood are in disagreement
    in their predictions about $\STFa$,
    with the shape likelihood predicting 
    a modest $\STFa > 0$.
This is an indication that we may need to re-examine
    our detection model.

In the above work,
    we have methodically examined our use of the Inhomogeneous
    Poisson Point Process to identify potential
    systematic bias in our model
    (for example, we consider potential bias in a mis-tuned detection model
    as well as an incomplete exploration of model parameter spaces).
However, the final analysis of our likelihood model
    does demonstrate a peak in a consistent region of parameter space.
Keeping in mind the sources of bias we have discussed,
    we characterize the preferred model parameters.
As seen in figure \ref{fig:D-series-joint-likelihood},
    we see a peak in likelihood at $\STwind=1$.
This is further explored by considering the three-dimensional 
    set of models where $\STwind=1$ is constant.
\correction{
The properties of the models which best match observation are given
    in table \ref{tab:D-series-stats},
    including our preferred model, D411.
The estimated uncertainty in our parameters for the models
    in the viscinity of parameter space with highest joint likelihood
    are generated by the Gaussian process,
    and regions of one- and two-sigma are observed in figure 
    \ref{fig:D-series-joint-likelihood}.
}

\end{subsection}
\end{section}

 \end{chapter}

\begin{chapter}{Conclusions}
\label{chap:conclude}
Throughout this dissertation I have introduced methods,
    demonstrated application, and shared our findings
    in modeling the properties of individual gravitational-wave events,
    and constraining population models for the isolated
    binary evolution formation channel for compact binaries.
In this final chapter, I provide a summary of our methods and results,
    and I present our conclusions.
Furthermore, I discuss the broader impact of my work on 
    parameter estimation, population synthesis, and the field of
    multi-messenger astronomy.

\begin{section}{Summary}

We present both compressed parametric and non-parametric methods
    for fully characterizing the astrophysical properties of
    each gravitational-wave event reported by modern
    gravitational-wave observatories.
We have demonstrated that our method is versatile and applicable
    to high-dimensional parameter spaces,
    including source-frame or detector-frame mass parameterizations,
    distance, precessing spin, and the tidal deformability of neutron stars.
We use these methods to also provide
    optimized parameterizations for the bounded (truncated)
    Normal Approximate Likelihood (NAL) models
    describing each event reported in the
    Gravitational-Wave Transient Catalogs.
Each NAL model has the added benefit of describing a maximum likelihood
    estimate for the properties of a given event
    which is unbiased by the finite boundary limitations
    imposed by similar models assumed from the sample mean.

We find that NAL models are in strong agreement with
    the parameter estimation sample density for events in the GWTC releases,
    even for a selection of sparsely simulated likelihood evaluations.
These models can be used to compare the properties of the gravitational-wave
    population to any population model (including popular Monte Carlo methods),
    due to their accuracy and computational efficiency.
Furthermore, their less biased maximum likelihood estimates for the parameters
    of each event can be synthesized to describe the bulk properties of
    the population on an event-by-event basis,
    such as the mass distribution and tendency 
    toward equal mass in compact binaries.
The characteristic $\sigma$ associated with the
    optimized truncated Gaussian also provides an excellent
    standard for describing the half-width-half-max of a truncated Gaussian
    and the significance of a deviation from
    equal mass and zero (or liminal) spin.
Our findings are also in agreement with collaborative works,
    in that populations of compact binaries observed by modern
    gravitational-wave detectors have mostly
    equal mass and zero spin
    (with notable exceptions).

Following this, we use the StarTrack binary evolution algorithm
    to construct synthetic universes representing a population of merging
    binaries under a range of assumptions about the underlying parameters
    of binary evolution.
Using these populations, we apply a detection model to predict 
    the rate of gravitational-wave observations
    and compare those predicted observations to those 
    reported by real gravitational-wave observatories. 
We apply Bayesian statistical inference
    to compare both the predicted detection rates and properties of predicted
    compact binaries to real observations reported by gravitational-wave
    observatories.
We extend our analysis using Gaussian process regression
    to consider the parameter space between simulations,
    fully characterizing a response
    in our range of allowed values for each formation parameter.
We provide our method for this analysis,
    as well as our conclusions about the impact of formation parameters
    on the isolated binary evolution formation channel.

With our simplified detection model,
    we demonstrate the effectiveness of our methods
    as we draw preliminary constraints on four formation parameters
    ($\STFa$, $\STbeta$, $\STkick$, and $\STwind$)
    for the isolated binary evolution formation channel.
Our analysis includes efficent techniques for comparing a large discrete set
    of population models to gravitational-wave observations,
    an interpolation model which can predict around those descrete models,
    as well as sytematic consistency tests which check
    if each component of our model is in agreement.
We find that both $\STFa$ and $\STwind$ have a strong impact on
    the gravitational-wave detection rate associated with a
    binary evolution model.
These parameters, together with the supernova engine assumed for
    the isolated binary evolution formation channel,
    must both be considered when analyzing the agreement
    of the detection rate predicted for a simulated merger population 
    to real observations.
Presently, our results suggest an unrestricted stellar mass loss due to wind,
    as well as substantial supernova recoil kicks,
    and significant mass and angular momentum loss due to ejected
    material during the common envelope phase of binary evolution
    (see sec \ref{sec:constraints-4D}).

\end{section}
\begin{section}{Broader Impacts of Science and Methodology}

The significance of this work has immediate consequences for
    related studies in the field of multi-messenger astronomy.
As many groups are interested in studying the formation channels
    of compact binary mergers,
    quickly assessing the agreement of sample mergers to the
    real gravitational-wave events reported in the GWTC releases
    is of immediate application.
NAL models provide this application and are available
    as a public release (\url{https://gitlab.com/xevra/nal-data}),
    together with our methods (\url{https://gitlab.com/xevra/gwalk}
    and \url{https://gitlab.com/xevra/gp-api}).
As gravitational-wave detectors move closer to design performance
    in future observing runs,
    inefficient sample-based estimates for
    the likelihood function of each gravitational-wave
    event for a population model will become increasingly computationally
    challenging.
NAL models can be used to fully characterize the astrophysical properties of
    each merger and can be evaluated in a small fraction of a second
    (even for high-dimensional parameter spaces),
    and will be ever more useful in upcoming observing runs
    for efficient population synthesis.

Conclusions about the formation parameters for the isolated binary evolution
    channel for the formation of compact binaries
    are also of immediate consequence to other work.
Our results can inform other population synthesis models of
    reasonable expectations for the physical processes in binary evolution,
    and what assumptions are valid to make.
More widely in multi-messenger astronomy,
    a simulated synthetic universe with predicted compact binary 
    detection rates in agreement with gravitational-wave observations
    will help to predict other physics by providing physically motivated
    sets of sample mergers representing stellar populations.
One example of these broader applications includes work done by my
    colleage, using our population synthesis model
    to constrain the neutron star equation of state \cite{holmbeck2021nuclear}.

Even directly,
    as our detection model, population of gravitational-wave signals,
    and simulation sample expand,
    we can further constrain our formation channel.
This will allow us to fit an even wider set of models
    for mass transfer in binary systems,
    mass loss due to stellar wind in stars
    with varying mass and metallicity,
    as well as supernova engines and ejecta.
Future work can further strengthen these constraints
    by considering the agreement of stellar populations predicted in
    these synthetic universes with multi-messenger observation,
    such as Gaia populations and electromagnetic neutron-star emission,
    and by considering dynamic mergers as other groups have proposed
    \cite{shikauchi2022detectablity}.
Such an analysis can fully apply the potential of multi-messenger astrophysics
    to form a more complete understanding of stellar populations and
    evolution in our Universe.

\end{section}

 \end{chapter}

\bibliographystyle{abbrv}
{\footnotesize\bibliography{Bibliography.bib}

\end{document}